\documentclass[12pt]{article}
\usepackage{amsmath}
\usepackage{enumerate}
\usepackage{url} 

\newcommand{\blind}{1}

\usepackage{graphicx}
\usepackage{amssymb, listings}
\usepackage{amsthm}
\usepackage{docmute}
\usepackage{comment}
\usepackage{mathtools}
\usepackage{bbm}
\usepackage{natbib}
\usepackage{algorithm}
\usepackage[noend]{algorithmic}
\newtheorem{theorem}{Theorem}
\newtheorem{lem}[theorem]{Lemma}
\newtheorem{prop}[theorem]{Proposition}

\newcommand{\hypo}{\mathrm{hypo}}
\newcommand{\conv}{\mathrm{conv}}
\newcommand{\conc}{\mathrm{conc}}
\newcommand{\dom}{\mathrm{dom}}
\newcommand{\cl}{\mathrm{cl}}
\newcommand{\argmax}{\mathop{\rm arg~max}}
\newcommand{\argmin}{\mathop{\rm arg~min}}
\theoremstyle{definition}
\newtheorem{example}[theorem]{Example}
\newtheorem*{remark}{Remark}
\pdfoutput=1

\begin{document}
\def\spacingset#1{\renewcommand{\baselinestretch}%
{#1}\small\normalsize} \spacingset{1}


\if1\blind
{
  \title{\bf Maximum Likelihood Estimation of Log-Concave Densities on Tree Space}
  \author{Yuki Takazawa\\
    Department of Mathematical Informatics, \\Graduate School of Information Science and Technology, \\University of Tokyo\\
    and \\
    Tomonari Sei \\
    Department of Mathematical Informatics, \\
      Graduate School of Information Science and Technology, \\University of Tokyo}
  \maketitle
} \fi

\if0\blind
{
  \bigskip
  \bigskip
  \bigskip
  \begin{center}
    {\LARGE\bf Maximum Likelihood Estimation of Log-Concave Densities on Tree Space}
\end{center}
  \medskip
} \fi

\bigskip
\begin{abstract}
	 Phylogenetic trees are key data objects in biology, and the method of phylogenetic reconstruction has been highly developed. The space of phylogenetic trees is a nonpositively curved metric space. Recently, statistical methods to analyze the set of trees on this space are being developed utilizing this property.
Meanwhile, in Euclidean space, the log-concave maximum likelihood method has emerged as a new nonparametric method for probability density estimation. In this paper, we derive a sufficient condition for the existence and uniqueness of the log-concave maximum likelihood estimator on tree space. We also propose an estimation algorithm for one and two dimensions.

Since various factors affect the inferred trees, it is difficult to specify the distribution of sample trees. The class of log-concave densities is nonparametric, and yet the estimation can be conducted by the maximum likelihood method without selecting hyperparameters. We compare the estimation performance with a previously developed kernel density estimator numerically. In our examples where the true density is log-concave, we demonstrate that our estimator has a smaller integrated squared error when the sample size is large. We also conduct numerical experiments of clustering using the Expectation-Maximization (EM) algorithm and compare the results with k-means++ clustering using Fr\'{e}chet mean.
\end{abstract}

\noindent%
\vfill

\newpage
\spacingset{1.5} 



\section{Introduction}
Inference of phylogeny is one of the key problems in biology. Phylogenetic trees represent the evolutionary histories of taxa. The methods of phylogenetic reconstruction using current gene data have been highly developed, and it is now possible to reconstruct them easily with various models and inference methods \citep{felsenstein2004inferring}.

However, due to the uncertainty of inference, incomplete lineage sorting or irregular biological processes such as horizontal transfer, it is usually the case that different gene loci indicate different evolutionary histories \citep{Reid2014}. The classical approach to this problem is finding a {\it consensus tree} \citep{bryant2003classification}, a single summary tree of multiple trees inferred from different loci, which is constructed by some rules. In 2001, \citet{Billera2001} introduced the space of phylogenetic trees with $n$ leaves, {\it tree space}, which is a geodesic metric space with nonpositive curvature. Recent research has shifted to the statistical analysis of a set of trees in such spaces, enabling a geometrical perspective. These efforts include point estimation by Fr\'{e}chet mean \citep{Benner2014}, principal component analysis \citep{Nye2011}, outlier detection using a kernel density estimator \citep{Weyenberg2014, Weyenberg2017}, and construction of confidence sets \citep{Willis2019}. 

Since numerous factors affect the inferred phylogenetic trees, the parametric approach to specifying the distribution of inferred trees is difficult, and the risk of misspecifying models is high. In this sense, the  nonparametric approach generally specifies fewer constraints in the distribution and might be desirable. The kernel density estimator proposed in \citet{Weyenberg2014, Weyenberg2017} is designed for this purpose.

In Euclidean space, log-concave density estimation has emerged as a new shape-constrained, nonparametric method of density estimation. Log-concave density is a class of probability densities whose logarithm are concave functions. Compared to classical smoothing approaches such as kernel density estimates, in which we need to specify some hyperparameters such as bandwidths, log-concave density has an advantageous property that the estimation can be done automatically. Concretely, it has been shown that the maximum likelihood estimators exist in both the one-dimensional and multi-dimensional cases and that the calculation of the estimators can be reduced to a convex optimization problem in $\mathbb{R}^n$ with $n$ sample points \citep{Cule2010b}. 

In this paper, we show that the maximum likelihood estimator of log-concave density in tree space exists under some conditions and that the estimation can be implemented in the one-dimensional case. Although we do not derive an algorithm for multidimensional cases due to the difficulty of finding the closure of convex hulls, we also show that we can approximately calculate the maximum likelihood estimator in the two-dimensional case.

The remaining sections are organized as follows. In section 2, we review some basic concepts of tree space and define some concepts for convex analysis in Hadamard spaces. Section 3 presents our main result for maximum likelihood estimation in tree space, in one dimension and multiple dimensions. Section 4 explains how to calculate the maximum likelihood estimator in the one-dimensional and two-dimensional cases. Section 5 shows the results of simulation studies. We compare the performance of density estimation with the kernel density estimator. We also compare the simulation results of clustering with a mixture of log-concave densities to the k-means++ approach using the Fr\'{e}chet mean.

\section{Preliminaries}
\subsection{Phylogenetic Trees and Tree Space}
A phylogenetic tree is modeled as a tree with labels only on the leaves. We call a tree with $n+1$ labeled leaves an $n$-tree. $n$-trees have one root leaf representing the common ancestor and $n$ leaves representing present taxa. Internal edges are edges that are not directly connected to leaves. In binary $n$-trees, it is easy to see that the number of internal edges is $n-2$. Nonbinary trees have fewer internal edges since they can be obtained by contracting some internal edges of some binary trees. The number of different topologies for binary $n$-trees is $(2n-3)!!$\citep{Biologists2008}. Note that the edges in different tree topologies can be regarded as the same if they partition the leaves in the same way. 
If each internal edge in an $n$-tree has a positive length, we call it a {\it metric $n$-tree}. 

\cite{Billera2001} modeled the space of metric $n$-trees as follows. First, each binary tree topology is modeled as a positive Euclidean orthant with dimension $n-2$, with each axis corresponding to the length of each internal edge of that topology. These orthants are stuck together on the axes representing the same edges. 
Nonbinary trees consitute the boundaries of these orthants since they are obtained by contracting some internal edges.
In particular, the $n$-tree without any internal edges is the point located at the center of tree space connected to every orthant. We call this point the {\it origin} in this paper. The space of metric $n$-trees obtained in this way is called {\it tree space} and is denoted as $\mathcal{T}_n$.

\begin{figure}[ht]
\begin{minipage}{0.32\linewidth}
	\centering
	\includegraphics[width=0.9\linewidth]{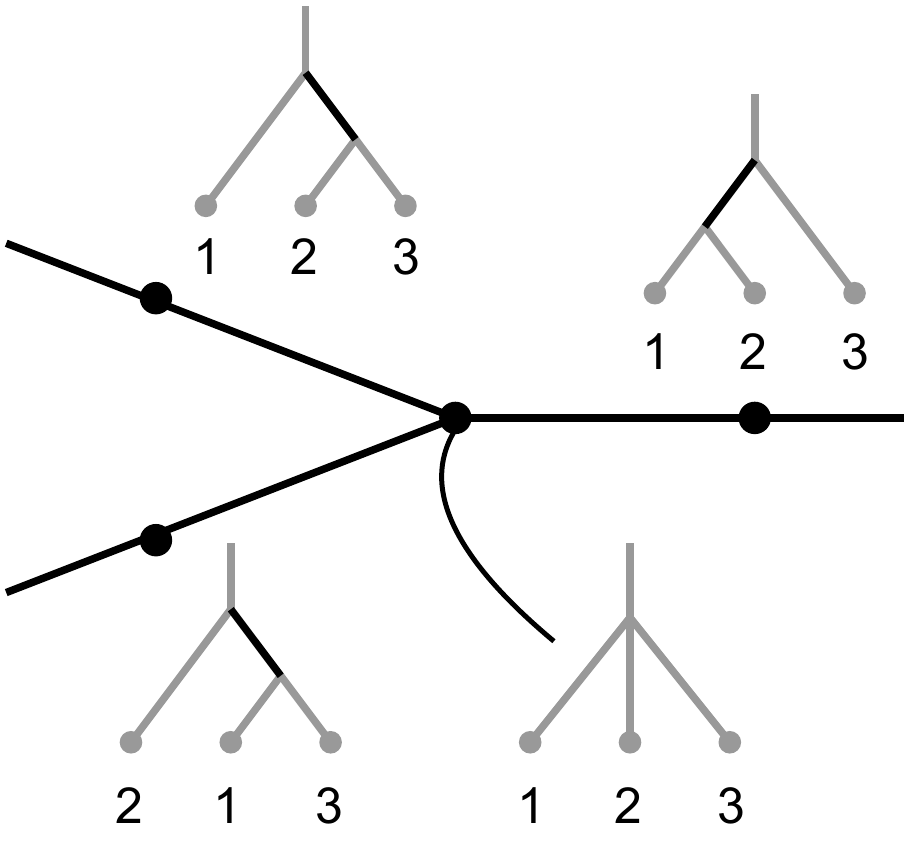}
\end{minipage}
\begin{minipage}{0.32\linewidth}
	\centering
	\includegraphics[width=0.9\linewidth]{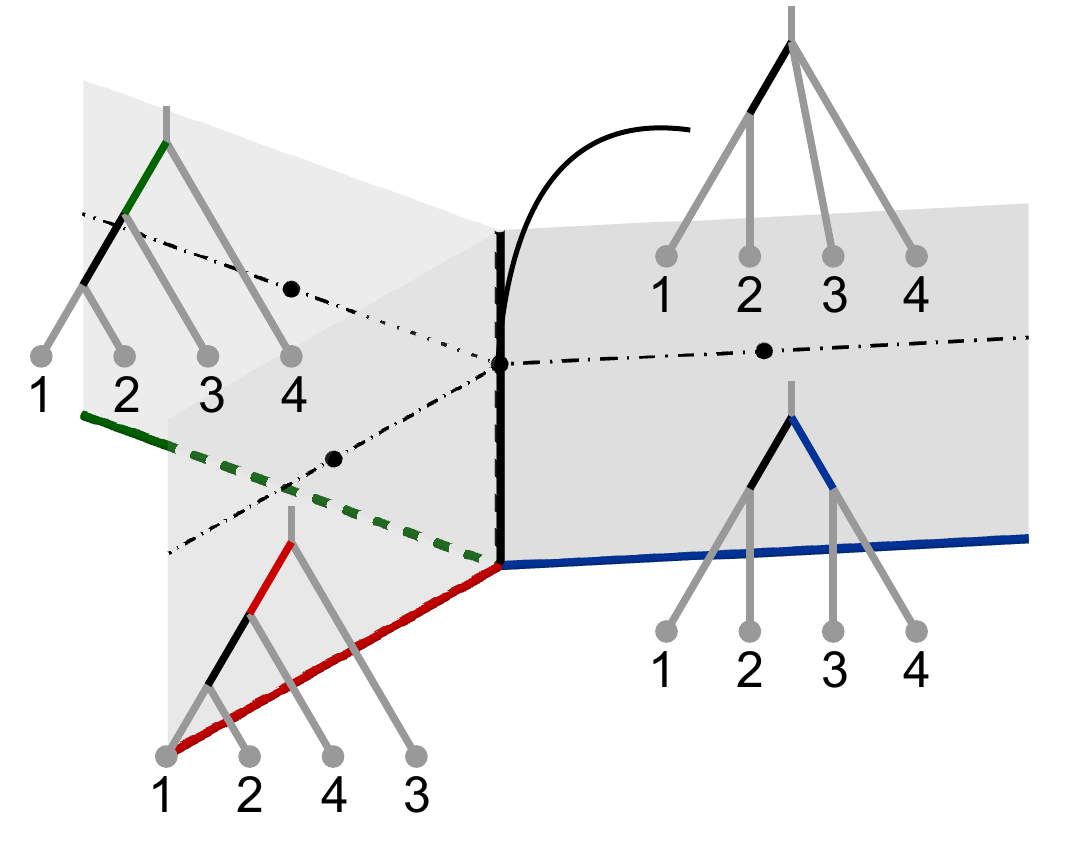}
\end{minipage}
\begin{minipage}{0.32\linewidth}
	\centering
	\includegraphics[width=0.9\linewidth]{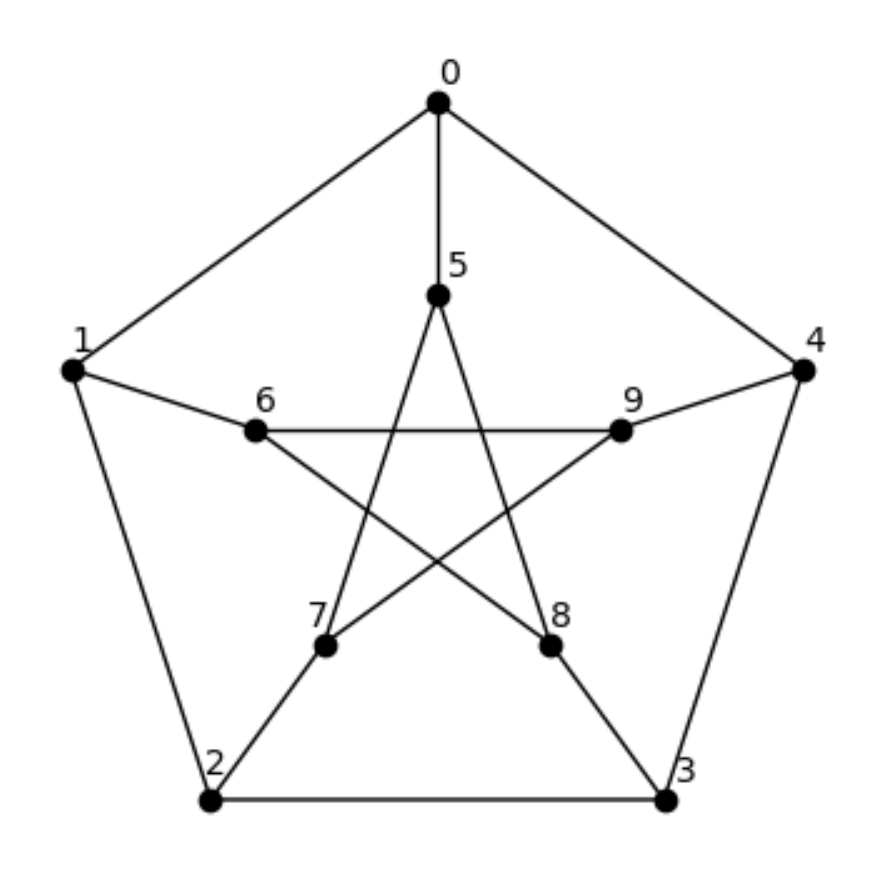}
\end{minipage}
\caption{Depiction of tree space. (Left): one-dimensional tree space $\mathcal{T}_3$, consisting of three half-lines each corresponding to a topology. (Center): part of two-dimensional tree space. (Right): Petersen graph. Each edge represents a 2-dimensional orthant in $\mathcal{T}_4$ with endpoints representing axes constituting the boundaries. Vertices corresponding to axes are indexed from $0$ to $9$ for convenience.  }
\label{figure_treespace}
\end{figure}

In this paper, we mainly discuss the results in the one-dimensional or two-dimensional case. By dimension, we mean the dimension of each orthant or the dimension of the space seen as a cubical complex. Therefore, by the $p$-dimensional tree space, we mean the space $\mathcal{T}_{p+2}$. In the one-dimensional tree space, we only have $3!! = 3$ topologies and each orthant representing some topology is a half-line connected to the other two at the origin, the point representing a trifurcating tree (Figure \ref{figure_treespace}). Two-dimensional tree space $\mathcal{T}_4$ has $5!! = 15$ topologies, each orthant being a nonnegative Euclidean plane, and three different orthants are connected to each axis (Figure \ref{figure_treespace}). It is known that if we represent each axis as a vertex and each orthant as an edge connecting two vertices, then the whole of space $\mathcal{T}_4$ can be graphically depicted as a Petersen graph, as shown in Figure \ref{figure_treespace}. For details, see \citet{Billera2001} or \citet{Lubiw2020}, for example.

\subsection{Geometry of Tree Space}
Tree space can be readily formulated as a metric space. Since it is the union of Euclidean orthants, we can simply define the distance between two points in the same orthant as the usual Euclidean distance. For any two points in distinct orthants, we can set the distance between them as the infimum of the lengths of paths with endpoints in those two points. The infimum is attained since tree space consists of a finite number of orthants, and once a sequence of orthants of a path is decided, there is a unique path with minimum length. The paths which attain the minimum distances are called {\it geodesics}, and a metric space with a geodesic between any two points is called a {\it geodesic metric space}. 

Tree space is also known to be a {\it Hadamard} space; i.e., it is a complete metric space with nonpositive curvature. See, for example, \citet{Bacak2014} for proof. 
This property is key to many important results in this space. In Hadamard spaces, the geodesic connecting any two points is guaranteed to be unique. \citet{OWEN2011} has developed an algorithm to find geodesics in tree space in $O(p^4)$ time, where $p$ denotes the dimension of tree space. In tree space, geodesics are piecewise segments through sequences of orthants, and in some cases, they become {\it cone paths}, where each geodesic consists of two segments connecting the endpoints and the origin. In particular, in two-dimensional tree space $\mathcal{T}_4$, whether a geodesic between two points becomes a cone path depends only on the ``angle'' between them. See, for example, \citet{Lubiw2020} for a detailed account of this property. The nonpositively curved property also results in the existence and uniqueness of the {\it Fr\'{e}chet mean}. The Fr\'{e}chet mean $\mu$ of given samples $X_1,\ldots,X_n\in\mathcal{H}$ and weights $w_1,\ldots, w_n$ is defined as the minimizer of the sum of squared distance:
\begin{align}
	\mu = \argmin_{x\in\mathcal{H}}\sum_{i=1}^n w_i d(x,X_i)^2.
\end{align}
It is a generalization of the arithmetic mean since they coincide with uniform weights in Hilbert spaces. It is known that the Fr\'{e}chet mean can be calculated using the proximal point algorithm \citep{Bacak2013}. See, for example, \citet{Bacak2014} for proof of results regarding the Fr\'{e}chet mean in Hadamard spaces.

\subsection{Convex Sets and Concave Functions on Hadamard Spaces}\label{convexanalysis}
In this section, we define several concepts of convex sets and concave functions on Hadamard spaces. 

First, we define convex sets as in \citet{Bacak2014}. Let $(\mathcal{H},d)$ be a Hadamard space and for any points $x,y \in \mathcal{H}$, let $[x,y]$ denote the unique geodesic between them. Also, for $\lambda\in[0,1]$, we use informally the notation $(1-\lambda) x + \lambda y$ to denote the points in the geodesic $[x,y]$ with distance $\lambda d(x,y)$ from $x$. We say $A\subseteq \mathcal{H}$ is convex if for any points $x,y\in A$ we have $[x,y] \subseteq A$. 

Function $f:\mathcal{H}\to[-\infty, \infty]$ is concave if its hypograph ${\hypo f} = \{(x,\mu) \in \mathcal{H} \times \mathbb{R}~|~ \mu\leq f(x)\}$ is convex. Note that a product of Hadamard spaces is also a Hadamard space, and thus, in particular, $\mathcal{H}\times \mathbb{R}$ is also a Hadamard space. An equivalent definition of concavity of a function $f$ can be given in the following way: given any points $x,y \in\mathcal{H}$, if $\alpha < f(x)$ and $\beta < f(y)$, then for any $\lambda\in(0,1)$, $f((1-\lambda)x+\lambda y) > (1-\lambda) \alpha + \lambda \beta$.

Given any set $S$ in $\mathcal{H}$, a convex hull of $S$ is the smallest convex set in $\mathcal{H}$ containing $S$. This set exists because the intersection of (possibly infinite) convex sets is convex and that $\mathcal{H}$ is convex. We denote the convex hull of $S$ as $\conv~ S$. In Hadamard spaces, it is known that a convex hull can be written as in the following Lemma.
\begin{lem}[\citet{Bacak2014}, Lemma 2.1.8]\label{lem:bacak1}
	For $S\subseteq \mathcal{H}$, put $C_0 = S$ and for $n\in\mathbb{N}$, recursively define $C_n$ by \\$C_n = \{x\in\mathcal{H}~|~ x \text{ is in a geodesic between points in }C_{n-1}\}$. Then, 
	\begin{align}
		\conv~ S = \bigcup_{n=0}^\infty C_n.
	\end{align}
\end{lem}

This Lemma indicates that taking geodesics between points an infinite number of times obtains the convex hull.
Next, we define an important concept of the concave hull of a function. Given any function $f$ on $\mathcal{H}$, the concave hull of $f$, $\conc~ f$, is defined as the least concave function minorized by $f$. Its existence and uniqueness in Euclidean space is a well-known result; see chapter 5 of \citet{rockafellar1970convex} for a detailed account. Here we show the existence and uniqueness of the Hadamard case, but as we will see, the proof proceeds in the same way as in the Euclidean case.

\begin{lem}\label{lemma:convhull}
	For any function $f:\mathcal{H}\to[-\infty, \infty]$, the concave hull $\conc~ f$ exists and is unique. 
\end{lem}
\begin{proof}
Let $f$ be an arbitrary function on $\mathcal{H}$, and define a function $g(x) = \sup\{\mu ~|~ (x,\mu) \in \conv(\hypo f)\}$. We show below that this $g$ is the desired concave hull. We first show that $g$ is concave. For any points $x,y\in\mathcal{H}$, if $\alpha < g(x)$ and $\beta < g(y)$, there exist $\varepsilon>0$ such that $\alpha+\varepsilon < g(x)$ and $\beta+\varepsilon < g(y)$. Then, $((1-\lambda)x + \lambda y, (1-\lambda)\alpha + \lambda \beta + \epsilon) \in \conv(\hypo~f)$ for any $\lambda\in(0,1)$. This leads to $g((1-\lambda)x + \lambda y) > (1-\lambda)\alpha + \lambda \beta$, thus showing that $g$ is indeed concave. Also, by definition, $g$ is minorized by $f$. The minimality of $g$ follows from the minimality of a convex hull of sets. The uniquness is obvious from the minimality.
\end{proof}

An important implication of the proof of the Lemma \ref{lem:bacak1} is how to derive the concave hull of a function: calculate the convex hull of its hypograph, and take the supremum. In the remaining sections, our interest is in cases where the function can be written in the following form:
\begin{align}
	f_y(x) = \begin{cases}
	    y_i &  \text{if } x=X_i, \\
	    -\infty & \text{otherwise}.
	\end{cases} \label{vertical_func}
\end{align} 
In this case, the hyopgraph of the function $f_y$ is the set of vertical half-lines:
\begin{align}
	\hypo f_y = \{(X_i, \mu)~|~\mu\leq y_i, i=1,\ldots, N\}. 
\end{align}

In the Euclidean case, the convex hull of this set is characterized by the convex combination of the $N$ points $X_1, \ldots, X_N$, resulting in the concave hull of $f$ being a piecewise linear function. However, in Hadamard spaces, the convex combination of $N$ points is not well defined in that associativity does not hold. What we know is that Lemma \ref{lem:bacak1} implies that any points $(x,\mu)$ in the convex hull of this set can be obtained by repeatedly taking the convex combination of two points. 
We use the notation $h_y = \conc f_y(x)$ to denote this concave hull. 
We refer to function $h_y$ as the {\it least concave function} with $h_y(X_i)\geq y_i~~(i=1, \ldots, n)$.

Also, note that as in the Euclidean case, the restrictions of convex (resp.\ concave) functions to a convex set are also convex (resp.\ concave). This is again because the intersection of two convex sets is also convex, so the hypograph of a restricted function is also convex. Following the usual convention in convex analysis, we regard a concave function $f$ that has as its domain some convex set $S\subseteq \mathcal{H}$ as a function on the whole of space $\mathcal{H}$ which takes $-\infty$ outside $S$. 

Another condition that is usually assumed with concavity is {\it upper-semicontinuity}. A function $f:\mathcal{H}\to[-\infty,\infty]$ is {\it upper-semicontinuous} at $x$ if $\limsup_{y\in\mathcal{H}:d(x,y)\to 0} f(y) \leq f(x)$. If function $f$ is upper-semicontinuous at all points $x\in\mathcal{H}$, then it is upper-semicontinuous on $\mathcal{H}$. It is also equivalent to the hypograph of $f$ being closed.

Similar to concave hulls, we can define an {\it upper-semicontinuous hull} of a function $f$ as the least upper-semicontinuous function minorized by $f$. 
That this function exists follows from the fact that (possibly infinite) intersections of closed sets are closed. Furthermore, it is possible to define the {\it upper-semicontinuous concave hull} of a function of $f$, as the least upper-semicontinuous concave function minorized by $f$. 
Concretely, this function is derived by taking intersections of the hypographs of all upper-semicontinuous and concave functions minorized by $f$.  It can also be characterized as the function whose hypograph is $\cl(\conv~\hypo f)$. We denote the upper-semicontinous concave hull of the function $f_y$ by $\bar{h}_y$.

\subsection{Probability Measure on Tree Space}
In this section, we introduce one of the possible constructions of a probability measure on tree space. This construction is due to \citet{Willis2019}. 

Let $(\mathcal{T}_{n}, d)$ be tree space. Note that this space consists of $(2n-3)!!$ Euclidean $(n-2)$-dimensional nonnegative orthants. Using this, one can write any set $A$ in $\mathcal{T}_{n}$ as a union of Euclidean sets, $A = \cup_{i=0}^{(2n-3)!!} A_i$. Here, $A_0$ denotes the set of trees in $A$ that are also in the boundary of $\mathcal{T}_n$, and $A_i$ denotes the set of trees in $A$ that are also in the $i$-th positive orthant. Now, let $\nu_B$ be the Lebesgue measure on $\mathbb{R}^{n-2}$, and define $\nu$ by $\nu(A) = \sum_{i=1}^{(2n-3)!!} \nu_B(A_i)$. Then this $\nu$ preserves sigma additivity, and by completion of measures, $\nu$ becomes a complete measure. 

In the remaining sections, we will use measure $\nu$ as a base measure and consider densities with respect to it.

\section{Existence and Uniqueness of Maximum Likelihood Estimator}
	\subsection{One-Dimensional Case}\label{subsec:onedim}
In this section, we consider the simplest tree space, $\mathcal{T}_3$. Let $\mathcal{F}_0$ be the set of log-concave probability densities with respect to the base measure $\nu$ on this space. 
The following theorem indicates that maximum likelihood estimation in $\mathcal{T}_3$ could be performed in the same manner as in the Euclidean case.

\begin{theorem}\label{theorem-1dim}
	Let $(X_1, \ldots, X_n)$ be an independent sample from $f\in\mathcal{F}_0$ in $\mathcal{T}_3$, with $n \geq 2$. Then, with probability 1,  the maximum likelihood estimator $\hat{f}_n$ of $f$ exists and is unique: i.e., $\hat{f_n}$ is the unique maximizer in $\mathcal{F}_0$ of the log-likelihood function $l(f) = \sum_{i=1}^n \log f(X_i)$. 
\end{theorem}

The proof is given in the supplementary material.

\begin{remark}
Although our main interest is in tree space, it is simple to see that the above argument also applies to the space of $N\in\mathbb{N}$ half-lines connected at the origin. 
\end{remark}
	\subsection{Multidimensional Case}
In this section, we consider the general $p$-dimensional tree space, $\mathcal{T}_{p+2}$. We let $\bar{\mathcal{F}_0}$ be the set of upper-semicontinuous log-concave densities in $\mathcal{T}_{p+2}$ and $l(f) = \sum_{i=1}^n \log f(X_i)$ denote the log-likelihood function. The maximum likelihood estimator in this case, unlike in the cases of the one-dimensional tree space and Euclidean space, might not even exist. Before deriving a sufficient condition for the existence of the MLE, we first show that when the maximizer exists, its uniqueness can be stated as in the following theorem. 

\begin{theorem}\label{theorem:ddim_uniqueness}
	Let $(X_1, \ldots, X_n)$ be a sample from some density on $\mathcal{T}_{p+2}$. Suppose that a maximizer of $l(f)$ in $\bar{\mathcal{F}_0}$ exists. Then the maximizer is unique $\nu-$almost everywhere.
\end{theorem}
\begin{proof}
	Suppose $f_1, f_2\in\bar{\mathcal{F}_0}$ both maximize $l(f)$. If we put \\ $f(x)=\{f_1(x)f_2(x)\}^{1/2}/\int_{\mathcal{T}_{p+2}}\{f_1(z)f_2(z)\}^{1/2}d\nu(z)$, $f$ is log-concave and
\begin{align}
	l(f) &= \frac{1}{2n} \sum_{i=1}^n \{\log f_1(X_i) + \log f_2(X_2)\} - \log\int_{\mathcal{T}_{p+2}}\{f_1(z)f_2(z)\}^{1/2}d\nu(z) \nonumber \\
	&\geq l(f_1) - \log\int_{\mathcal{T}_{p+2}}\frac{f_1(z)+f_2(z)}{2}d\nu(z) = l(f_1). 
\end{align}

Here, the last inequality is the relationship between the arithmetic mean and geometric mean, and equality holds if and only if $f_1(z) = f_2(z)~ \nu\mathrm{-a.e.}$.
\end{proof}

The next theorem provides a sufficient condition for sample points such that the maximum likelihood estimator exists and is unique almost everywhere.

\begin{theorem}\label{ddim_result}
	Let $(X_1, \ldots, X_n)$ be a sample from a density $f$ on $\mathcal{T}_{p+2}$, $n\geq p+1$. Let $C_n = \conv \{X_1, \ldots, X_n\}$, and  suppose that in each nonnegative orthant $\{O_i\}_{i=1}^{(2p+1)!!}$, one of the following conditions holds:
	\begin{enumerate}[(a)]
		\item $\cl(C_n)\cap O_i$ does not contain any points.
		\item $\cl(C_n)\cap O_i$ constitutes a $p$-dimensional set in $O_i$: i.e., $\nu(C_n\cap O_i) > 0$. \label{ddim}
		\item $\cl(C_n)\cap O_i$ only includes points at the boundaries, say $B_j~~(j=1, \ldots, J)$, and each $B_j$ is connected to at least one orthant $O_j$ in which (\ref{ddim}) holds.
	\end{enumerate}
	Then, a maximum likelihood estimator $\hat{f_n}$ of $f$ in $\bar{\mathcal{F}_0}$ exists: i.e., $\hat{f_n}\in\bar{\mathcal{F}_0}$ is a maximizer of the log-likelihood function $l(f) = \sum_{i=1}^n \log f(X_i)$. Moreover, it is unique except in some set of $\nu$-measure zero excluding $\mathrm{int}(C_n)$.
\end{theorem}

Note that the sufficient condition is a condition that holds with a probability approaching one as $n$ goes to infinity. Also, observe that this condition is easy to check if the convex hull of samples in $\mathcal{T}_{p+2}$ can be calculated.

The proof is given in the supplementary material. In the proof of Theorem \ref{theorem-1dim} and Theorem \ref{ddim_result}, it is shown that the maximum likelihood estimator of $n$ samples are parametrized by $n$-dimensional vector $y\in\mathbb{R}^n$. Concretely, the logarithm of the maximum likelihood estimator $\hat{f_n}$ is written in the form $\bar{h}_y$, the least upper-semicontinous concave function with $h_{y}(X_i)\geq y_i$. Note that in one-dimensional case, the least concave function is upper-semicontinuous; i.e. $h_y = \bar{h}_y$. It is also shown that we need to maximize the objective function $\psi_n(y)$ defined as follows:
\begin{align}
    \psi_n(y) = n^{-1}\sum_{i=1}^n \bar{h}_y(X_i) - \int_{\mathcal{T}_{p+2}}\exp(\bar{h}_y)(x)d\nu(x).
\end{align}

We also see in the proof that in some cases, the addition of a new point outside the existing convex hull does not induce an increase in the measure of the convex hull of the new set of points. Next is an example.

\begin{figure}[ht]
	\centering
		\includegraphics[width=0.6\linewidth]{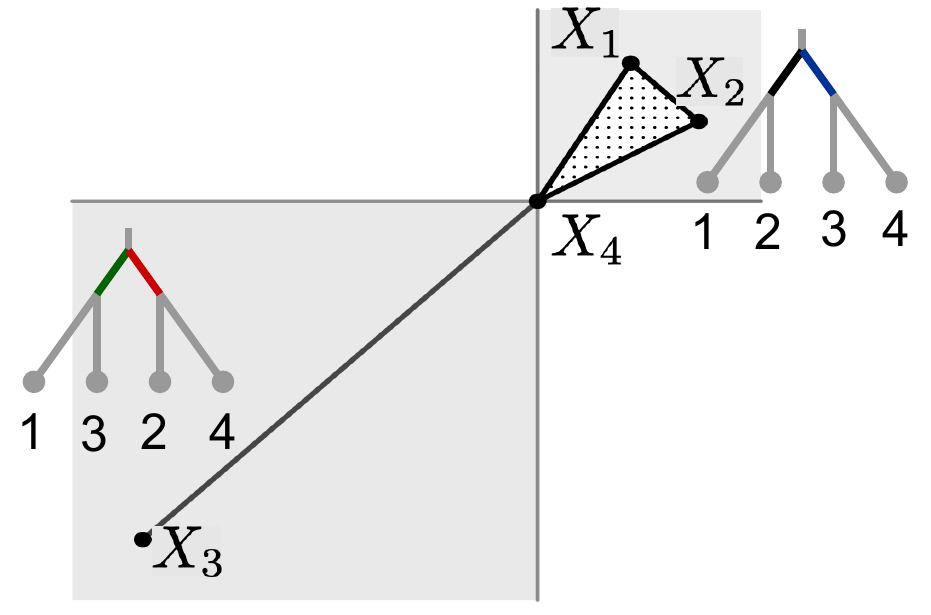}
		\caption{Illustration of Example \ref{example:no_change_nu} and Example \ref{no_unique_mle}. The two trees depicted are the topologies that the two orthants represent. Geodesics between these orthants are always cone paths. }
		\label{nomlefig}
\end{figure}

\begin{example}[nonuniqueness]\label{example:no_change_nu}
	Consider that we have three points $X_1, X_2, X_4 \in \mathcal{T}_4$ as depicted in Figure \ref{nomlefig}. Then $\conv\{X_1, X_2, X_4\}$ is the triangle $X_1X_2X_4$. Suppose that we add $X_3$ and consider $\conv\{X_1, X_2, X_4, X_3\}$. Then this set is the union of triangle $X_1X_2X_4$ and the segment $X_4X_3$, and it is clear that the measure has not increased: i.e.,  $\nu(\conv\{X_1, X_2, X_4\}) = \nu(\conv\{X_1, X_2, X_4, X_3\})$. Suppose that we have obtained  $f\in\bar{\mathcal{F}_0}$ that satisfies $f(x)>0~(x\in\conv\{X_1, X_2, X_4\}), f(x)=0~ (x\not\in\conv\{X_1, X_2, X_4\})$. Then ``extending'' the support of $f$ to include the segment $X_3X_4$ preserving log-concavity would not change the value of the likelihood function, and the extension of $f$ would also be a density. Thus, $\nu$-almost sure uniqueness holds, but strict uniqueness does not hold in this case.
\end{example}

Cases in which the maximum likelihood estimator does not exist occur when the convex hull has a measure-zero intersection with some orthant. 

\begin{example}[nonexistence]\label{no_unique_mle}
Consider that we have three samples $X_1, X_2, X_3 \in \mathcal{T}_4$ as depicted in Figure \ref{nomlefig}. The geodesics between points in these two trees necessarily are cone paths. It is clear that the assumption in Theorem \ref{ddim_result} is not satisfied. Now, assume that the lengths of segments $X_1X_4$ and $X_2X_4$ are the same, and the length of segment $X_3X_4$ is three times that of $X_1X_4$ and $X_2X_4$. Let $\log f$ be an affine-like function with $\log f(X_3) = y+\Delta$ and $\log f(X_2) = \log f(X_1) = y-\Delta/3$. Then $\log f(X_4) = y$ and \begin{align}\psi_n(f) = y +\frac{\Delta}{9} - \int_{C} f(x)dx.\end{align}

One can see that as $\Delta$ goes to $+\infty$ the second term linearly goes to $+\infty$ and the last integral term converges to zero. Thus, $\psi_n$ is a monotone increasing function with respect to $\Delta$ and does not attain a maximum. Therefore, the maximum likelihood estimator does not exist in this case. 
\end{example}

\subsection{Densities That Bend at the Boundary}\label{sec:bend}

Although the nonparametric class of log-concave densities seem to be large enough to model various densities, there are densities that arise narturally, and are slightly not log-concave. In this section, we focus on the simplest one-dimensional tree space $\mathcal{T}_3$, and discuss a way to incorporate densities that are not log-concave at the boundary. Let $O_1,O_2,O_3$ represent the three half-lines, and let the restriction of a density $f:\mathcal{T}_3\to\mathbb{R}$ to each half-line $O_i$ be $f_i$, which is a function on $\mathbb{R}_{\geq 0}$. Denote the exterior derivative at the origin by $\partial f_i / \partial n_i(0)$, defined as:
\begin{align}
    \frac{\partial f_i}{\partial n_i}(0) = \lim_{h\to +0} \frac{f_i(0)-f_i(h)}{h}.
\end{align}

\cite{Nye2014b} considered diffusion process on $\mathcal{T}_3$ and defined a natural generalization of Gaussian density as a solution to the heat equation. If diffusion starts from the point $x_0\in\mathbb{R}_{> 0}$ in the orthant $O_1$, the probabilty density at time $t$ in each orthant can be written as follows:
\begin{align}
\begin{split}
f_1(x) &= \phi(x;x_0,t) - \frac{1}{3}\phi(x;-x_0,t) \\
f_j(x) &= \frac{2}{3}\phi(x;-x_0,t) ~~~~~ (j\in\{2,3\})
\end{split} \label{browndensity}
\end{align}
Here, $\phi(x;\mu,\sigma^2)$ denotes the probability density function of normal distribution on $\mathbb{R}$ with mean $\mu$ and variance $\sigma^2$.

Density \eqref{browndensity} does not have log-concavity since it ``bends'' at the origin. In fact, the exterior derivatives are 
\begin{align}
\begin{split}
     \frac{\partial f_1}{\partial n_1}(0) &= -\frac{4}{3}\phi^\prime(0;x_0,t), \\
     \frac{\partial f_j}{\partial n_j}(0) &= \frac{2}{3}\phi^\prime(0;x_0,t)~~~~~ (j\in\{2,3\}).
\end{split}
\end{align} 
This makes density $f$ to be not log-concave around the origin.

Another example can be constructed by considering a simple coalescent process of three lineages from three different species.
\begin{example}\label{ex:coalescent}
    Consider the species tree drawn on the left of Figure \ref{fig:coalescent}, and that three lineages are drawn from three species. In a neutral coalescent process \citep{KINGMAN1982235} with $n$ lineages, the distribution of time that takes for any two lineages to coalesce is given by simple exponential distribution:
    \begin{align}
        p(t) = \binom{n}{2}\exp\left({-\binom{n}{2}t}\right)
    \end{align}
    
    In a multispecies coalescent \citep{rannala2003bayes}, one assume that the two lineages coalesce only after the corresponding two species coalesce. Let $T$ be the time where exactly two clades of species exist as shown in the Figure \ref{fig:coalescent}. We can write the distribution of the length of internal edges by
    \begin{align}
        f_i(x) = \begin{cases}
        -\frac{1}{6}\exp({-x-T}) + \frac{1}{2}\exp\left({-x+T - 2\max(0,T-x)}\right) & \text{if }i=1\\
        \frac{1}{3}\exp({-T-x}) & \text{if }i\in\{2,3\},
        \end{cases}
    \end{align}
    where $i$ denotes an index of orthant. See the supplementary material for the derivation.
    
    This density is illustrated in Figure \ref{fig:coalescent} for the case $T=1$. We can clearly see that it is not log-concave. In fact, the exterior derivative is
    \begin{align}
        \frac{\partial f_i}{\partial n_i}(0) = 
        \begin{cases}
            -\frac{2}{3}\exp(-T) & \text{if }i=1\\
            \frac{1}{3}\exp(-T) & \text{if }i\in\{2,3\}\\
        \end{cases}
    \end{align}
    
    \begin{figure}
    \begin{minipage}{0.32\linewidth}
        \centering
        \includegraphics[width=.95\linewidth]{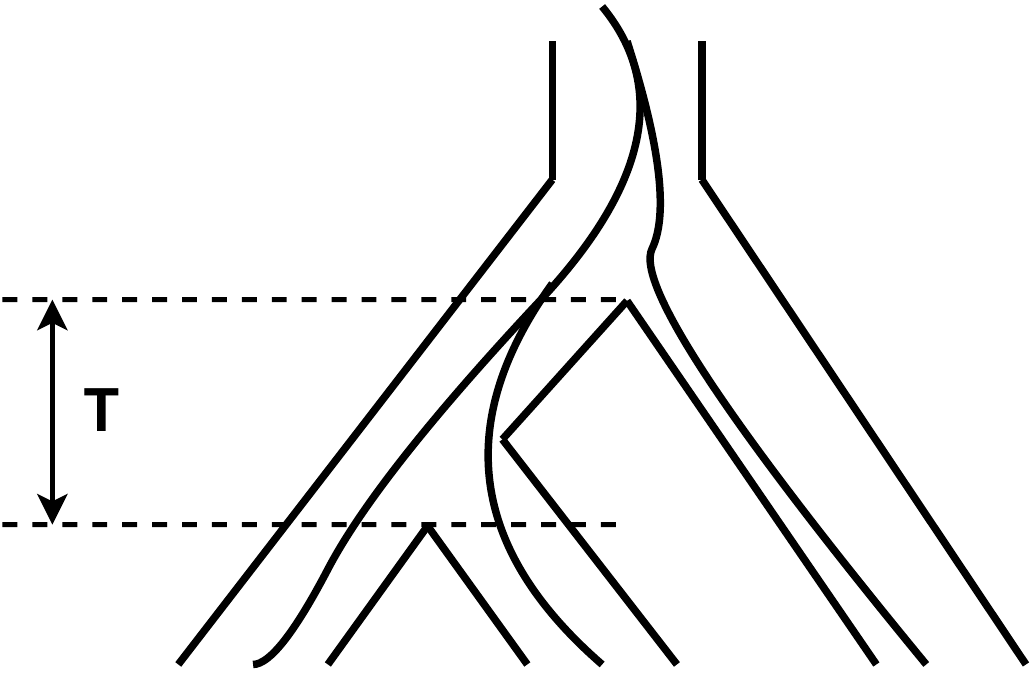}
    \end{minipage}
    \begin{minipage}{0.32\linewidth}
        \centering
        \includegraphics[width=\linewidth]{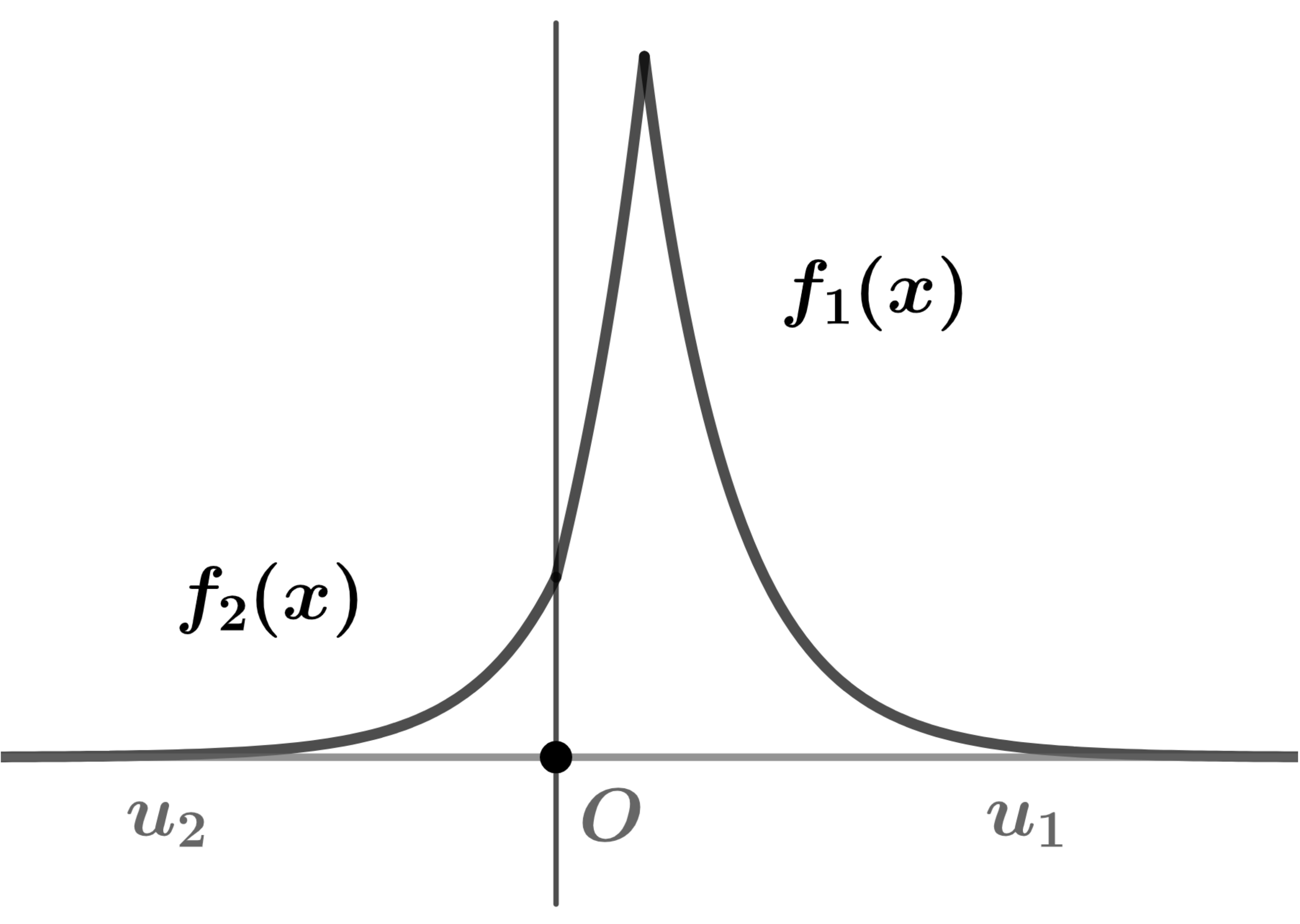}
    \end{minipage}
    \begin{minipage}{0.32\linewidth}
        \centering
        \includegraphics[width=\linewidth]{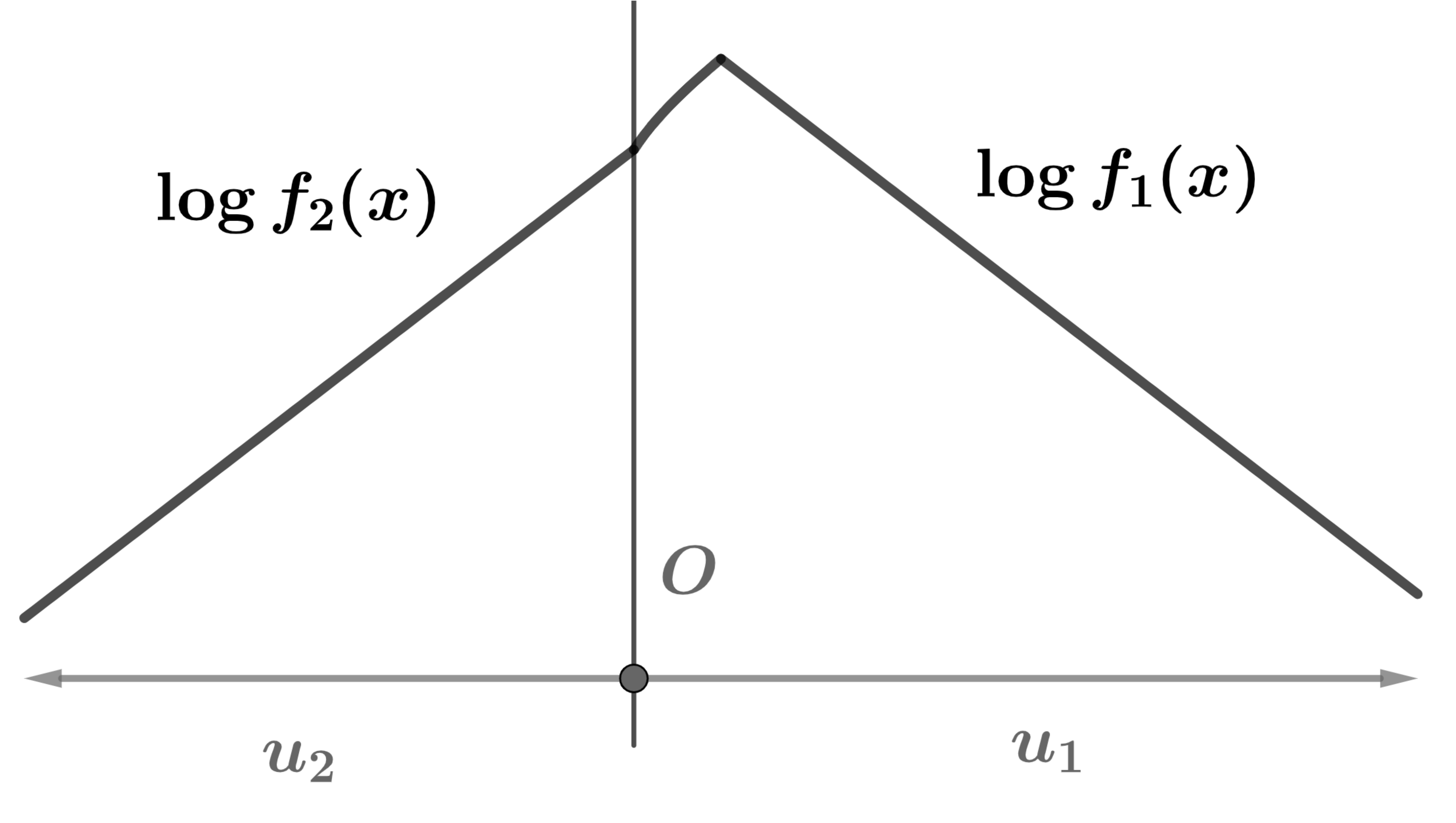}
    \end{minipage}
    \caption{(Left): An example of how three lineages inside a species tree coalesce. $T$ represents the length of internal edge in the species tree. (Center): The probability density in the orthants $O_1$ and the orthant $O_2$. (Right): The logarithm of the probability density in the orthants $O_1$ and $O_2$.}
        \label{fig:coalescent}
    \end{figure}
    
\end{example}

In these two examples, densities restricted to each half-line $\mathbb{R}_{\geq 0}$ are log-concave. The only non log-concave part is at the origin, where the absolute value of the slope of density differs. Concretely, the following relationship holds:
 \begin{align}
        \frac{\partial f_1}{\partial n_1}(0) \leq 0, ~~~~~\frac{\partial f_1}{\partial n_1}(0) = -2\frac{\partial f_{j}}{\partial n_{j}}(0)~~~~~j\in\{2,3\}.
    \end{align}
We can also see that it satisfies the following Kirchhoff-type condition: 
\begin{align}
    \sum_{i=1}^3 \frac{\partial f_i}{\partial n_i}(0) = 0.
\end{align}

These densties that bend at the origin can be incorporated in the model. We relax convexity condition of $\hypo(\log f)$ to be the following:

\begin{enumerate}
    \item $\hypo(\log f)\cap O_i\times\mathbb{R}$ is convex for all $i\in\{1,2,3\}$ 
    \item For $x_1\in O_i$ and $x_2\in O_j$ with $i\neq j$, let $\lambda=d(x_1,0)/d(x_1,x_2)$, and\\ \begin{align}y_0 = \min\left\{\frac{2(1-\lambda)\log f(x_1) + \lambda \log f(x_2)}{2-\lambda}, \frac{ (1-\lambda)\log f(x_1) + 2\lambda \log f(x_2)}{1+\lambda}\right\}.\end{align} Then, $(0,y_0)\in \hypo(\log f)$
\end{enumerate}

We denote by $\mathcal{G}_0$ the class of densities which satisfy these two conditions.
Any log-concave density $f$ satisfies these two conditions since
$y_0 \leq (1-\lambda)f(x_1) + \lambda f(x_2)$.
Also, this allows for the bending at the origin to certain extent. In fact, the two densities we considered in this section is included in $\mathcal{G}_0$.

We can now exploit similar arguments to the one-dimensional log-concave case in section \ref{subsec:onedim} and derive the existence and uniqueness of the maximum likelihood estimator in $\mathcal{G}_0$.
Note that the resulting estimator does not require Kirchhoff-type condition.

\section{Calculation of Maximum Likelihood Estimator in One and Two Dimensions}

For the maximization of the function $\psi_y$, we need to be able to calculate $h_y$ for a given input vector $y$.
In the one-dimensional case, this probem is reduced to the calculation of the convex hull of points $\{(X_1, y_1), \ldots, (X_n, y_n)\}$. We explain below that this can be reduced to convex hull calculation in $\mathbb{R}^2$, and the algorithm to calculate the log-concave maximum likelihood estimator is implementable. 

In the two-dimensional case, the calculation of $\bar{h}_y$ requires an algorithm to find the closure of the convex hull of points $\{(X_1, y_1), \ldots, (X_n, y_n)\}$ in $\mathcal{T}_4\times \mathbb{R}$. It is not known if we can obtain such algorithm, but we show below that obtaining an approximate convex hull is possible.

\subsection{Calculation of the Least Concave Functions on $\mathcal{T}_3$}\label{onedimcalc}
In order to obtain the maximum likelihood estimator (MLE), we need to be able to calculate the concave hull $h_y$ of a function of the following type:
\begin{align}
	f_y(x) = \begin{cases}
	    y_i &  \text{if } x=X_i, \\
	    -\infty & \text{otherwise}.
	\end{cases} 
\end{align} 
As discussed in section \ref{convexanalysis}, concave hulls can be found by first taking the convex hull of the hypograph and then taking the supremum. In this case, the problem is to find the convex hull $D_n$ of a finite set of sample points in $\mathcal{T}_3\times\mathbb{R}$. 

$\mathcal{T}_3$ is a space consisting of three half-lines $\mathbb{R}_{\geq 0}$ connected at the origin. If all sample points are from one orthant, then the situation is the same as the Euclidean case, so we consider the case that the sample points are from at least two orthants. 
Let $y_0 = \max\{ (1-\lambda)y_i + \lambda y_j~|~ i,j\in\{1,\ldots,n\}, \lambda\in[0,1], \gamma_{X_i,X_j}(\lambda) = 0\}$. Then, $(0,y_0)$ needs to be in the convex hull $D_n$. Furthermore, if we take convex hull of sample points in each orthant includeing $(0,y_0)$ and take their union, this set, say $E_n$, becomes a convex set. Obviously $E_n$ is included in $D_n$, thus by minimality of convex hulls, this set $E_n$ in fact equals $D_n$. Since $y_0$ can be calculated by taking geodesics of all pairs of sample points, one can calculate $D_n$ easily using the convex hull algorithm in Euclidean space $\mathbb{R}^2$. 

Similar arguments hold if we allow for the bend at the origin as discussed in the section \ref{sec:bend}. Concretely, we only need to alter the definition of $y_0$ to be 
\begin{align}
    y_0 = \max\left\{ \min\left\{\frac{\lambda y_j + 2(1-\lambda)y_i}{2-\lambda}, \frac{2\lambda y_j + (1-\lambda)y_i}{1+\lambda}\right\}~\middle|~ i < j, \lambda\in[0,1], \gamma_{X_i,X_j}(\lambda) = 0\right\}.\nonumber
\end{align}

\subsection{Approximation of $\bar{h}_y$ on $\mathcal{T}_4$} \label{2d_convhull}
In the two-dimensional case, it is not even known whether the convex hull of a finite number of points is closed. Recently, \cite{Lubiw2020} constructed an algorithm to find the closure of the convex hull of a finite number of points in single-vertex 2D CAT(0) complexes, including tree space, utilizing linear programming. Although we cannot transfer the algorithm directly to the product space of $\mathcal{T}_4$ and $\mathbb{R}$, \cite{Lubiw2020} also showed that by iteratively updating the values at the boundaries, one can construct a sequence of sets that converge to the desired convex hull. With a slight modification to this argument, as we show below, we can also create a similar situation so that by stopping at some iteration number, we can obtain an approximation of the convex hull.

To see this, let $S_0$ be all the sample points $\{(X_i, y_i)\mid X_i\in\mathcal{T}_4, y_i\in\mathbb{R}\}$. For simplicity, we will only consider the situation where the origin is in the convex hull (if the convex hull of sample points does not contain the origin, then the situation is the same as the Euclidean case). We call a geodesic in the space $\mathcal{T}_4 \times \mathbb{R}$ a cone path if it crosses the point of the type $\{0, y_l\}$, where $0$ denotes the origin in $\mathcal{T}_4$: that is, we call a geodesic a cone path when its projection onto $\mathcal{T}_4$ is a cone path. Also, for any nonnegative orthant $O$ in $\mathcal{T}_4$, define $S_0(O)=S_0\cap (O\times\mathbb{R})$, $H_0(O) = \conv(S_0(O))$ and $H_0 = \cup_{O} H_0(O)$. Following the terminology of \cite{Lubiw2020}, we define the sets $S_l\subseteq \mathcal{T}_4\times\mathbb{R}~(l=0,1,2,\ldots)$ (which \cite{Lubiw2020} call  the {\it $l$-th skeletons}) and $H_l\subseteq \mathcal{T}_4\times\mathbb{R}~(l=0,1,2,\ldots)$ iteratively as follows. 
 

First, we take geodesics between points in $H_{l-1}$ that are cone paths and find the values at the origin, which can be represented as $(0,y_{l,k})$. Then we find maximum and minimum values of $\{y_{l,k}\}$, and let them be $y_{l1}, y_{l0}$, respectively. We initialize $T_l$ with $S_{l-1}\cup \{(0, y_{l1}), (0, y_{l0})\}$. Although it is practically difficult to find the exact maximum and minimum values, we can use an approximation algorithm. We discuss on this approximation in the supplementary material. 

Now for each pair of points in $T_l$, take the geodesic between them, and add all intersection points with the boundaries to $T_l$. Here, by a boundary, we mean a two-dimensional plane formed by an axis on $\mathcal{T}_4$ and $\mathbb{R}$. Then, for each boundary in $\mathcal{T}_4$, take all points that are also in $T_l$, including points at the origin, which are of the form $(0,y_{li})$. Since each boundary is a two-dimensional plane, these points can be thought of as points in $\mathbb{R}^2$, and we take the usual two-dimensional Euclidean convex hull in this space. Then discard any points that are not vertices of this convex hull from $T_l$, and define $S_l = T_l$. Further, let $O$ be any nonnegative orthant in $\mathcal{T}_4$ and $S_l(O)$ be the points in $S_l$ that are also in $O\times\mathbb{R}$. Define $H_l(O) = \conv(S_l(O))$, which can be computed easily since $S_l(O)$ can be embedded into a nonnegative orthant in $\mathbb{R}^3$, and $H_l = \cup H_l(O)$. 

This construction is similar to the two-dimensional CAT(0) case \citep{Lubiw2020}, but the differences are that in this case, we have to determine the maximum and minimum values at the origin first, and instead of preserving only extreme points in a one-dimensional axis, we need here to save all the vertices in the convex hull of points on boundaries. 

To validate the first step, we first show the following lemma. 
\begin{lem}
	If $S_{l-1}$ is included in $\conv(S_0)$, then so is $S_l$.
\end{lem}
\begin{proof}
$S_{l-1} \subseteq \conv(S_0)$ indicates $H_{l-1}\subseteq \conv(S_{0})$, and thus $\conv(H_{l-1})\subseteq \conv(S_{0})$. By construction, $S_l \subseteq \conv(H_{l-1}) \subseteq \conv(S_{0})$.
\end{proof}
The next theorem ensures that $H_l$ converges to $\conv(S_0)$.
\begin{theorem}\label{convhull2d}
	Assume $p,q\in H_{l-1}$. Then the geodesic from $p$ to $q$, denoted as $\gamma_{p,q}$, is included in $H_l$. This indicates that $\cup H_l = \conv(S_0)$.
\end{theorem}

\begin{proof}
	If $\gamma_{p,q}$ is a cone path, let $(0, y_{0pq})$ be the point at which the geodesic crosses the origin. By construction, $S_l$ includes points $(0, y_u)$ and $(0, y_l)$ such that $y_l\leq y_{0pq} \leq y_u$. Therefore, $H_l$ includes $(0, y_{0pq})$, and thus all points of $\gamma_{p,q}$ are included in $H_l$.
	
	If $\gamma_{p,q}$ is not a cone path, then the simplest case is that $p,q$ are in the same orthant. In such a case, by construction, $\gamma_{p,q}\subseteq H_{l-1}$ as well. Thus, in particular, $\gamma_{p,q}\subseteq H_{l}$. Otherwise, let $p=(p_x,p_y), p_x\in\mathcal{T}_4, p_y\in\mathbb{R}$, and $q=(q_x,q_y), q_x\in\mathcal{T}_4, q_y\in\mathbb{R}$. We can take two nonnegative orthants $O_p, O_q\in\mathcal{T}_4$ to which $p_x$ and $q_x$ belong in such a manner that $O_p$ and $O_q$ are connected with at most one another orthant between them. Here, note that three connected orthants can be embedded in a part of $\mathbb{R}^2$ in the manner shown on the left of Figure \ref{EuclideanEmbedding}. Now, we can take $a_p, b_p, c_p \in S_{l-1}(O_p)$ having the following property: the triangle $a_pb_pc_p$ includes the points of the form $(p_x,y_{pp})$ with $y_{pp}\geq p_y$. These points are the ones that form the triangle ``over'' point $p$. In the same way, we can take $a_q, b_q, c_q\in S_{l-1}(C_q)$ having the property that the triangle $a_qb_qc_q$ includes the points of the form $(q_x,y_{qq})$ with $y_{qq}\geq p_y$. The case when the triangles exactly include $p$ and $q$ are drawn in Figure \ref{EuclideanEmbedding}.
	Let the intersection points of $\gamma_{pq}$ with one of the boundaries, say $B$, be $r=(x_r, y_r)$. Also, in the Euclidean embedding of $O_p$, $O_q$, and $B$, as in Figure \ref{EuclideanEmbedding}, let $-B$ denote the axis in the direction opposite to $B$ (Figure \ref{EuclideanEmbedding}). Recall that a ``boundary'' here means the Euclidean plane an axis in $\mathcal{T}_4$ makes in the space $\mathcal{T}_4\times\mathbb{R}$. Consider embedded space $\mathbb{R}^3$ which includes $O_p$ and $O_q$. If we allow paths to cross the part of the Euclidean space in which no orthants are embedded (the hatched region in Figure \ref{EuclideanEmbedding}), it is easy to see from Euclidean geometry that we can find two straight lines between points in $\{a_p,b_p,c_p\}$ and $\{a_q,b_q,c_q\}$, say $\gamma_1$ and $\gamma_2$, such that the line segment between the intersection points of $\gamma_1$ and $\gamma_2$ with $B$ or $-B$, say $r_1 = (x_{r1}, y_{r1})$ and $r_2 = (x_{r2}, y_{r2})$, includes a point of the form $(x_r, y_r^\prime)$ with $y_r^\prime\geq y_r$. If both $r_1$ and $r_2$ lie in $B$, then we just showed that $(x_r, y_r^\prime) \in H_{l}$. If $r_1$ or $r_2$ does not lie in $B$, say $r_1$ does not, then we still find that the segment from $r_1$ to $r_2$ goes ``under'' $(0,y_{l1})$, since if we let $(0,y_{r_1r_2})$ be the origin point included in the segment $r_1r_2$, then this point is included in some cone path between a point in the triangle $a_pb_pc_p$ and a point in the triangle $a_qb_qc_q$. Thus, the segment from $(0,y_{l1})$ to $r_2$ goes ``over'' $(x_r, y_r^\prime)$: i.e., the segment includes a point of the form $(x_r, y_r^{\prime\prime})$ with $ y_r^{\prime\prime}\geq y_r^\prime \geq y_r$. Thus in summary, by setting either $\tilde{y_r}=y_r^\prime$ or $\tilde{y_r}=y_r^{\prime\prime}$, we are able to find a point $(x_r, \tilde{y_r})\in H_l$ such that $\tilde{y_r}\geq y_r$. A similar argument leads to that we can find a point $(x_r, \hat{y_r})\in H_l$ such that $\hat{y_r}\leq y_r$. This indicates $r=(x_r,y_r)\in\mathcal{H}_l$, and consequently, the whole $\gamma_{pq}$. 
\begin{figure}\label{apbpcp}
	\begin{minipage}{0.25\linewidth}
	\begin{center}
		\includegraphics[width=\linewidth]{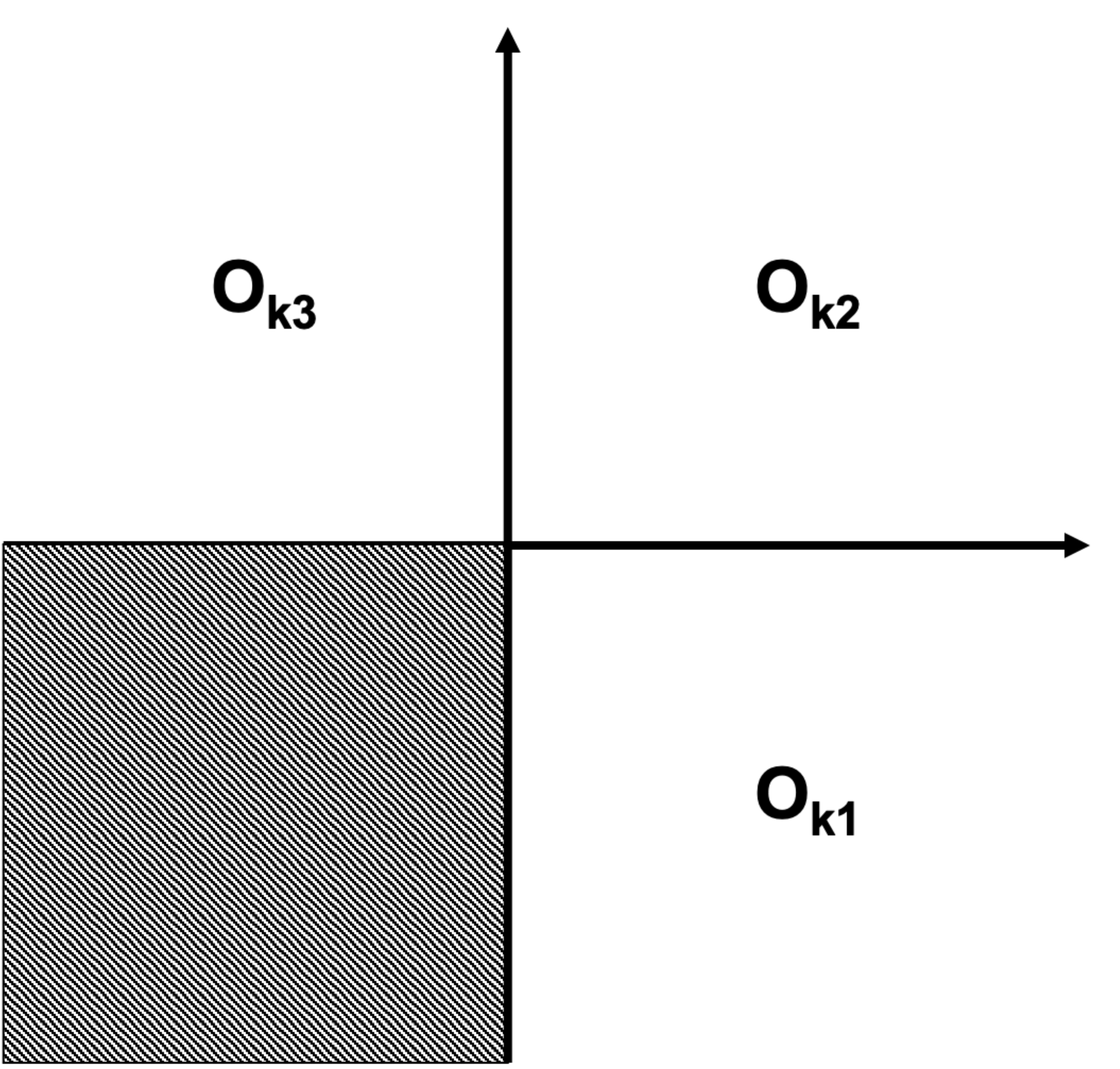}
	\end{center}
	\end{minipage}
	\begin{minipage}{0.37\linewidth}
	\begin{center}
		\includegraphics[width=\linewidth]{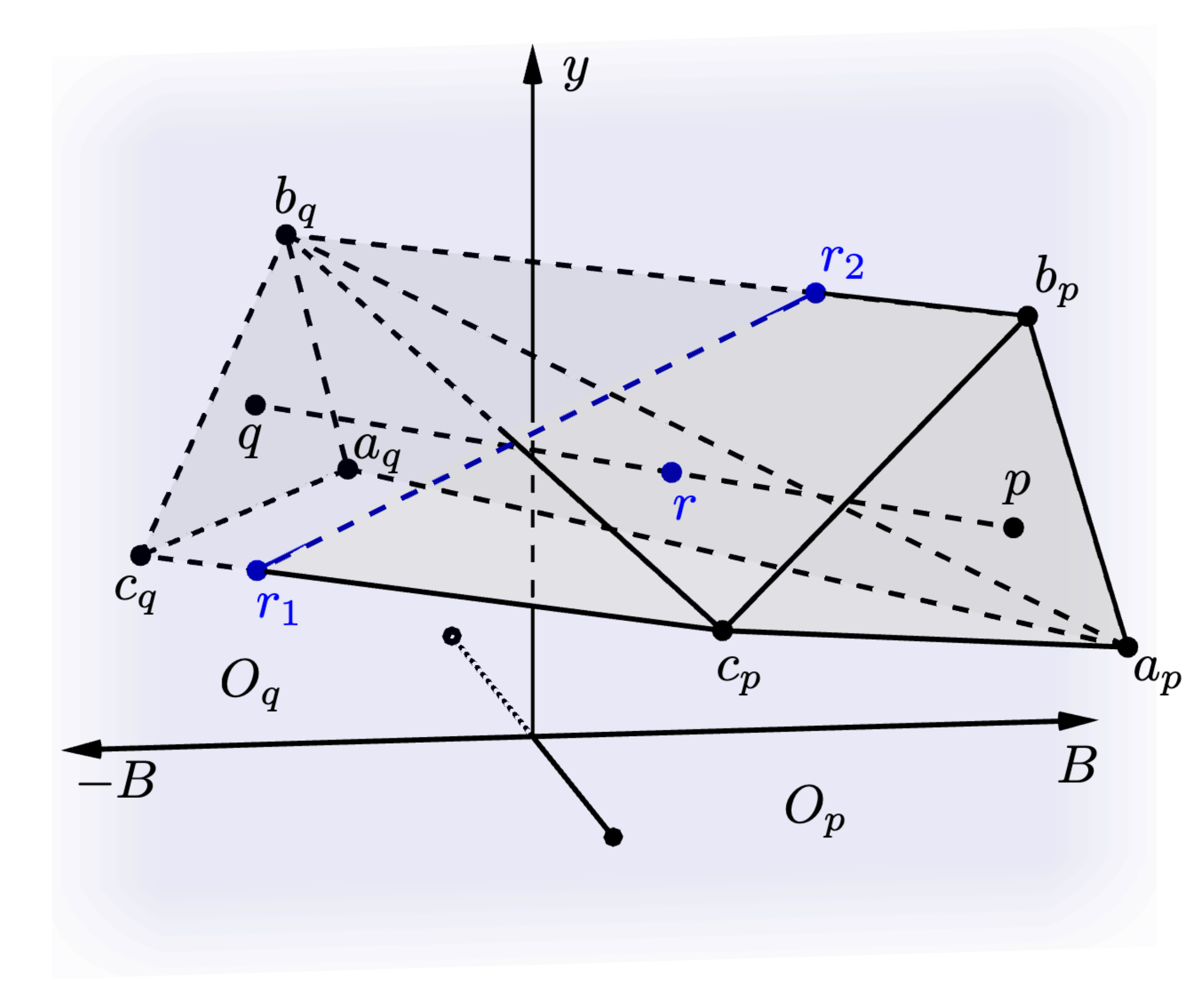}
	\end{center}
	\end{minipage}
		\begin{minipage}{0.37\linewidth}
	\begin{center}
		\includegraphics[width=\linewidth]{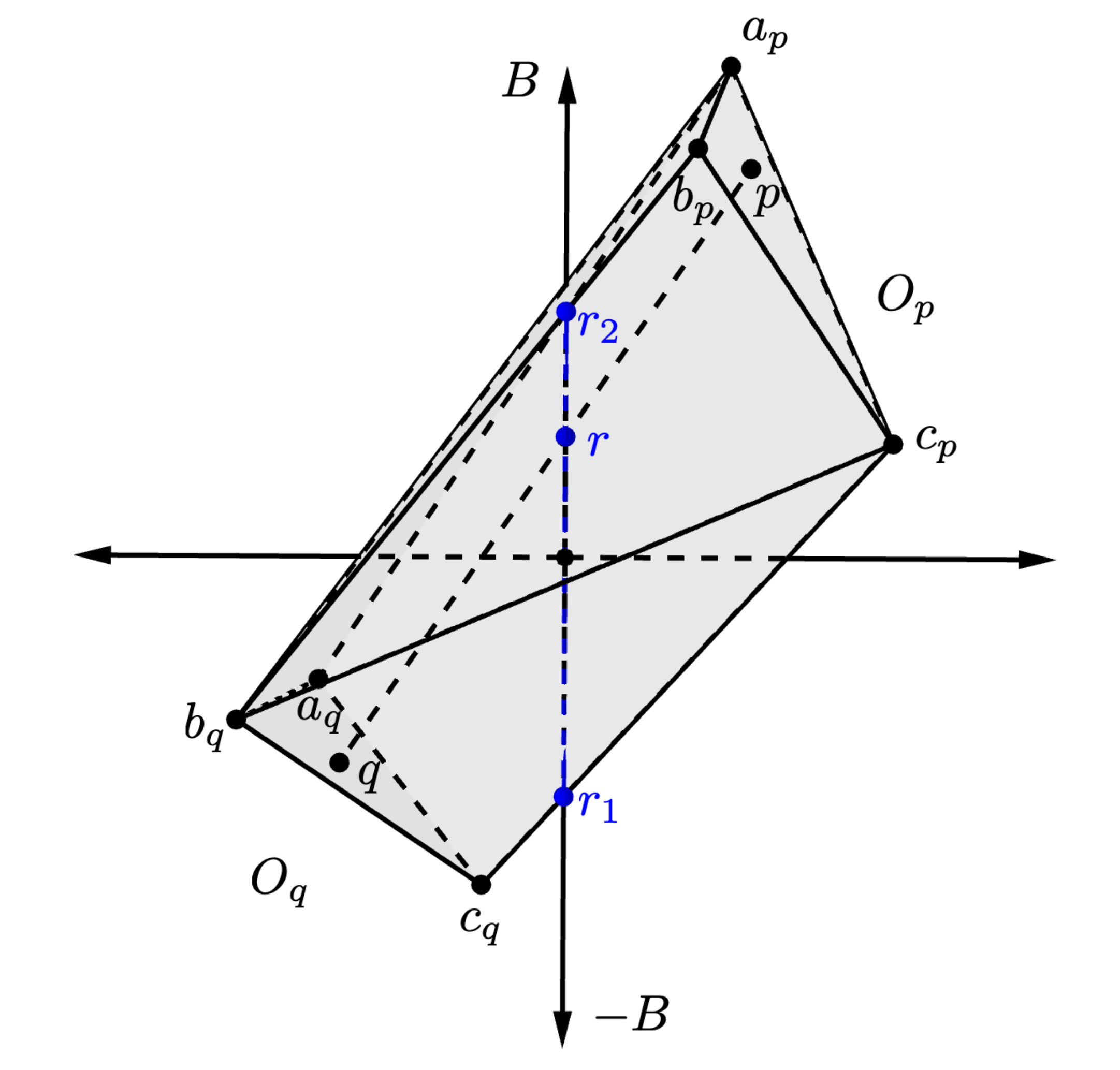}
	\end{center}
	\end{minipage}
	\caption{(Left): embeddings of three connected orthants to the Euclidean space $\mathbb{R}^2$. (Center): an illustration that the path $pq$ goes under the segment $r_1r_2$. The three blue points are included in the blue plane. (Right): central figure seen from the direction of $y$-axis.}
		\label{EuclideanEmbedding}
\end{figure}

\end{proof}

In \citet{Lubiw2020}, a similar fact was revealed, and they proceed to reduce the problem of finding the closure of a convex hull to a linear programming problem. However, such a reduction is difficult in our case, given that we do not have prior knowledge about how many vertices $\conv(S_0)$ would have on each boundary. We will content ourselves here with the approximation algorithm derived from the previous theorem. Note that if we can find maximum and minimum values at the origin exactly in the first step and the sets $S_l$ converge in finite iteration number $m$, the set $H_m$ is exactly equal to the desired convex hull. Furthermore, $H_m$ is closed in this case, thus function $\bar{h}_y$ calculated by this set is the exact least upper-semicontinuous concave function.

\subsection{Altering the Objective Function}\label{altering} Sections \ref{onedimcalc} and \ref{2d_convhull} show that we can calculate the function $\bar{h}_y$ for given $y$ in one-dimensional and two-dimensional cases at least approximately. As shown at the beginning of this section, in order to find the maximum likelihood estimator, we need to solve the optimization problem of the form $\max_{y\in\mathbb{R}^n} \psi(\exp(h_y))$. As in the Euclidean case \citep{Cule2010b}, the objective function is not a convex function of $y$, but by properly altering it, we can make this a convex optimization problem.
In the $d$-dimension case, the new objective function $\sigma_n(y)$ is defined as 
\begin{align}
	\sigma_n(y) = -\frac{1}{n}\sum_{i=1}^n y_i + \int_{\mathcal{T}_{d+2}} \exp(h_y(x))d\nu(x). \label{altered_objective}
\end{align}
The convexity of $\bar{h}_y$ with respect to $y$ can be derived in exactly the same manner as in the Euclidean case \citep{Cule2010b}. This ensures that $\sigma$ is a convex function of $y$. Therefore, we can utilize ordinary solvers for convex optimization to calculate MLE.

\section{Numerical Study}
In this section, we give two simulation results. First, we show that the log-concave distribution can be estimated properly in one and two dimensions. We also test the estimation of one-dimensional density that bends at the origin. Then we compare this result with the kernel density estimator proposed by \citep{Weyenberg2014}. Finally, we give an example of clustering with the log-concave mixture model.
\subsection{Estimation of One-Dimensional Log-Concave Densities}
For one-dimensional examples, we consider the following two types of log-concave densities, using $O_i~(i=1,2,3)$ to denote the three orthants in $\mathcal{T}_3$.

\begin{figure}[ht]
	\begin{center}
	\includegraphics[width=.9\linewidth]{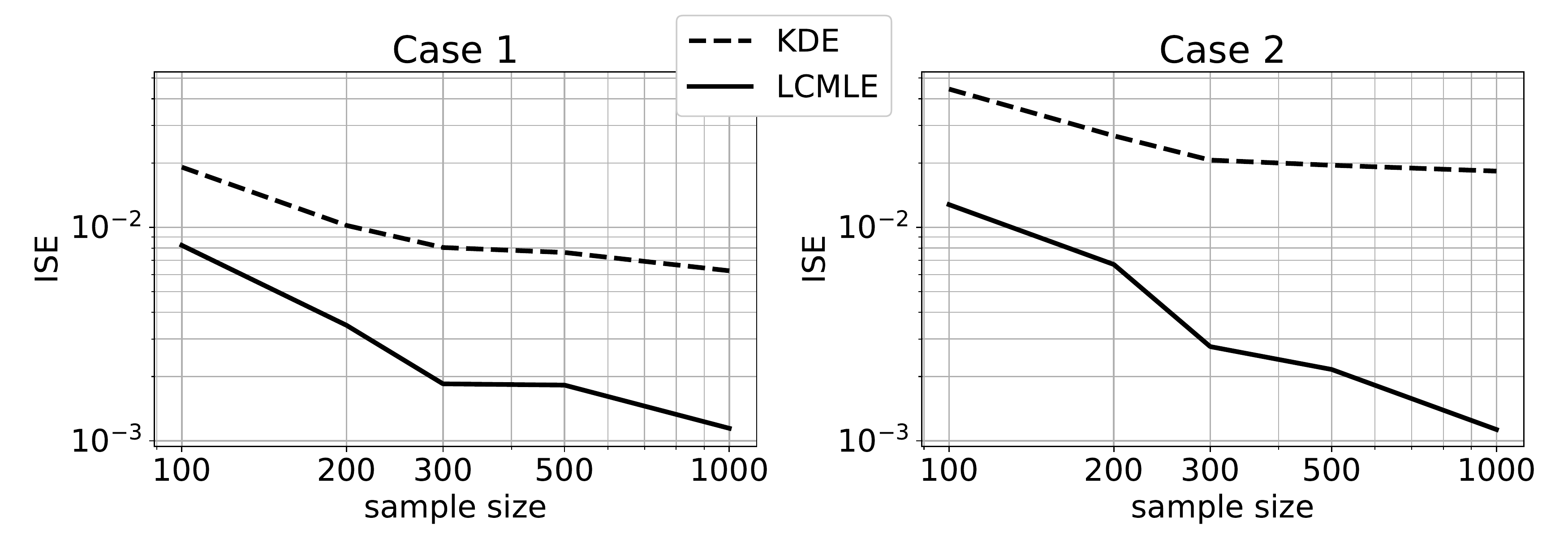}
	\end{center}
	\caption{Average ISE for sample sizes 100, 200, 300, 500, and 1000 in case 1 (left) and case 2 (right). LCMLE denotes log-concave MLE, and KDE denotes kernel density estimator.}
	\label{fig_est2}
\end{figure}

\begin{enumerate}
\renewcommand{\labelenumi}{Case \arabic{enumi}.}
	\item Normal-like density $f_1$ with mean equal to the point in $O_1$ that is 1 unit from the origin, $x_0$: i.e., for a point $x\in\mathcal{T}_3$ , $f_1(x) \propto\exp(-d(x,x_0)^2/2)$.
	\item Exponential-like density $f_2$, defined as follows:
	\begin{align}
		f_2(x) \propto \begin{cases} \exp(-d(x,x_0)) & \text{if }x\in O_2\cup O_3 \text{ or } (x\in O_1 \text{ and } d(x,0)<1) \\
		0 & \text{otherwise} .
		\end{cases}
	\end{align} 
\end{enumerate}
We generated ten samples each for sample size 100, 200, 300, 500, and 1000 from the true density and calculated the integrated squared error (ISE). The average ISE versus sample size is reported in Figure \ref{fig_est2}. 

Our simulation results show that the log-concave MLE dominates the kernel density estimator for all sample sizes. 

\subsection{Estimation of Two-Dimensional Normal-like Densities}
We denote each orthant of $\mathcal{T}_4$ by the indices of the axes at the boundary as in Figure \ref{figure_treespace}. For instance, the orthant with axes $0$ and $1$ is denoted by $\{0,1\}$. 
We consider densities of the following types:
\begin{enumerate}
\setcounter{enumi}{2}
\renewcommand{\labelenumi}{Case \arabic{enumi}.}
	\item Positive truncated multivariate normal density $f_3$ with mean 0 and variance $I$, supported on all orthants in $\mathcal{T}_4$:
	\begin{align}f_3(x) \propto \exp(-d(x,0)^2/2).\end{align}
	\item Positive truncated multivariate normal density $f_4$ with mean 0 and variance $I$, supported on orthants\\ 
	$\{0,1\}, \{1,6\}, \{6,8\}, \{3,8\}, \{3,4\}, \{0,4\}$:
	\begin{align}
		f_4(x) \propto 
		\begin{cases}
			\exp(-d(x,0)^2/2) & \text{if $x$ is in one of the above orthants} \\
			0 & \text{otherwise.}
		\end{cases}
	\end{align}
\end{enumerate}
Also, because of the nonpositive curvature property of tree space, we have that these densities are log-concave. 

\begin{figure}[ht]
    \begin{center}
        \includegraphics[width=.9\linewidth]{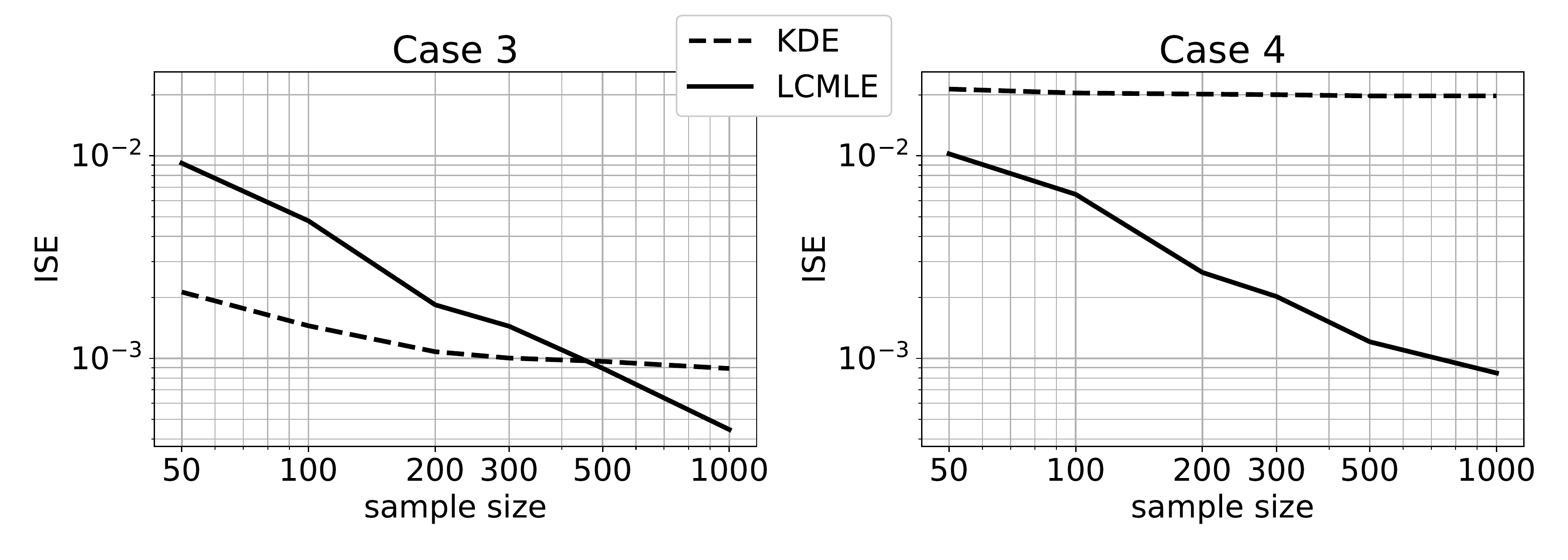}
    \end{center}
	\caption{(Left) average ISE in case 3. (Right) average ISE in case 4. LCMLE denotes log-concave MLE, and KDE denotes kernel density estimator.}
	\label{fig_ISE}
\end{figure}

We compare our log-concave MLE to the kernel density estimator \citep{Weyenberg2014} in terms of ISE with respect to the true density. 
We generated ten samples each for sample size 50, 100, 200, 300, 500 and 1000 from the true density and calculated MLE and the average ISE. The estimation results are illustrated in Figure \ref{fig_ISE}.


In case 3, the kernel density estimator performs better in small samples, but when a large number of samples are available, log-concave MLE starts to outperform the kernel density estimator. This result is similar to the Euclidean case \citep{Cule2010b}.
On the other hand, in case 4, log-concave MLE dominates the kernel density estimator. This is because while log-concave MLE can find the support of the true density with enough data, the kernel density estimator cannot. 

\subsection{Estimation of One-Dimensional Densities That Bend at the Origin}
As one-dimensional densities that bend at the origin, we take two densities we considered in the section \ref{sec:bend}. Concretely, we consider the following two densities:

\begin{enumerate}
\setcounter{enumi}{4}
\renewcommand{\labelenumi}{Case \arabic{enumi}.}
	\item Normal density constructed by Brownian motion $f_5$ at time 5 and starting position equal to the point in $O_1$ that is 1 unit away from the origin:
	    \begin{align}
f_{5,1}(x) &= \phi(x;1,5) - \frac{1}{3}\phi(x;-1,5) \\
f_{5,j}(x) &= \frac{2}{3}\phi(x;-1,5) ~~~~~ (j\in\{2,3\})
\end{align}

    \item Density of coalescent time in Example \ref{ex:coalescent} $f_6$ with $T=1$:
	i.e., for a point $x\in\mathcal{T}_3$ ,
   \begin{align}
        f_{6,1}(x) &= 
        -\frac{1}{6}\exp(-x-1) + \frac{1}{2}\exp\{-x+1 - 2\max(0,1-x)\} \\
        f_{6,j}(x) &= \frac{1}{3}\exp(-x-1) ~~~~~j\in\{2,3\}
    \end{align}
\end{enumerate}

\begin{figure}[ht]
	\begin{center}
	\includegraphics[width=.9\linewidth]{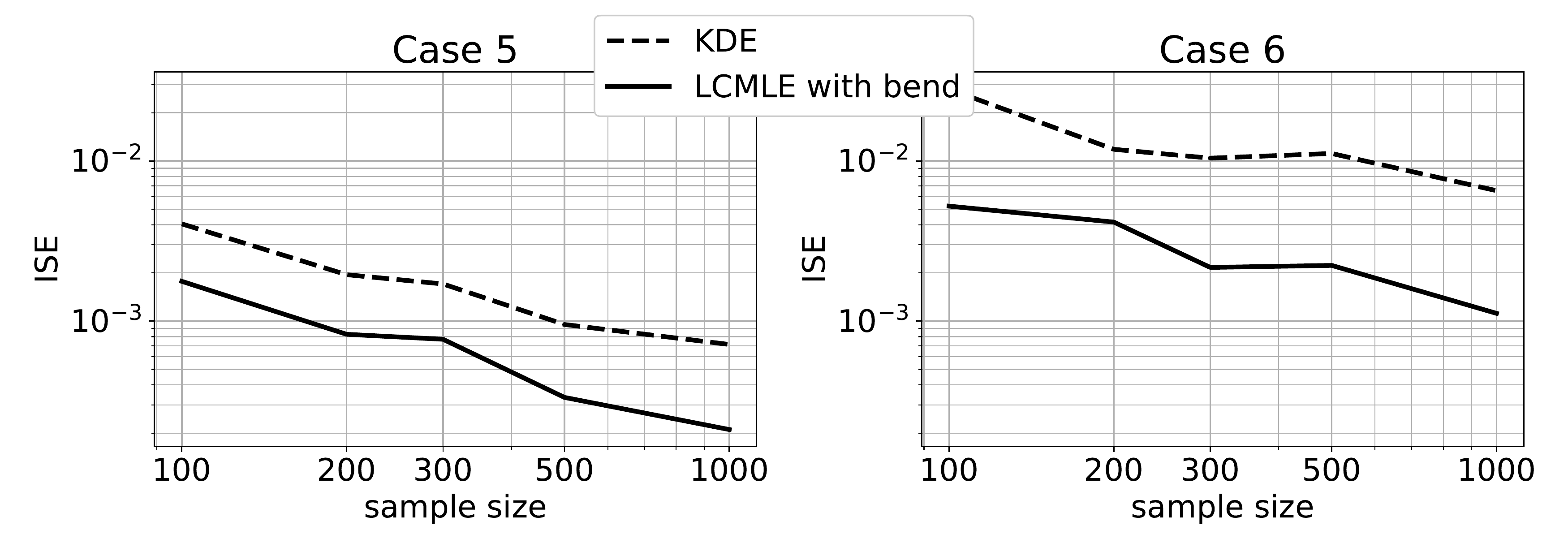}
	\end{center}
	\caption{Average ISE for sample sizes 100, 200, 300, 500, and 1000 in case 5 (left) and case 6 (right). LCMLE with bend denotes MLE in $\mathcal{G}_0$, and KDE denotes kernel density estimator.}
	\label{fig_estbend}
\end{figure}

As in the previous experiment with one-dimensional log-concave density, we generated ten samples each for sample size 100, 200, 300, 500 and 1000 from the true density and calculated MLE and the average ISE. The estimation results are illustrated in Figure \ref{fig_estbend}.

The maximum likelihood estimator in $\mathcal{G}_0$ dominated the kernel density estimator for all sample sizes.

\subsection{Clustering Example}
In this section, we consider the problem of clustering in the space $\mathcal{T}_4$. Let $x_1$ be a point in the orthant $\{0,1\}$ and $x_2$ be a point in the orthant $\{2,3\}$, both located at coordinates $(1,1)$ in their orthant. We consider two normal-like densities on $\mathcal{T}_4$, $g_1$ and $g_2$, defined as follows:
\begin{align}
	g_1(x) \propto \exp(-d(x,x_1)^2/2), \\
	g_2(x) \propto \exp(-2d(x,x_2)^2).
\end{align}
Note that $g_1$ and $g_2$ differ in variance as well as mean. 
We generated data from the mixture of $g_1$ and $g_2$, $f=\pi_1 g_1 + (1-\pi_1) g_2$ with proportion $\pi=0.5$, and we attempted to cluster these points using log-concave MLE.

As in the Gaussian mixture in Euclidean space, we can construct an Expectation-Maximization algorithm (EM algorithm) to solve the clustering problem. We will assume that we know the number of clusters to be $K=2$. We model the density in the following equation:
\begin{align}
	\hat{f}=\hat{\pi_1} \hat{g_1} + (1-\hat{\pi_1}) \hat{g_2},
\end{align}
where $\hat{g_1}$ and $\hat{g_2}$ are assumed to be log-concave. \citet{Cule2010b} explained the EM algorithm applied to a log-concave density, and we adopt their framework here. Briefly, given data points $X_i$, the algorithm proceeds in the following way in the general case with $K$ clusters. Given the current estimates of the mixture proportion and densities, $\hat{\pi_k}, \hat{g_k}$, the expectation step can be computed as 
\begin{align}
	P(Z_i=k\mid X_i) = \frac{\hat{\pi_k} \hat{g_k}(X_i)}{\sum_{j=1}^K \hat{\pi_j} \hat{g_j}(X_i)},
\end{align}
where $Z_i\in\{1,2,\ldots,K\}$ are the latent variables indicating the cluster the data come from. 
The maximization step is divided into two parts. First, for the estimate of $\hat{f}$, we maximize
\begin{align}
	\sum_{i=1}^n P(Z_i=k\mid X_i) \log g_k(X_i) \label{weighted_likelihood}
\end{align}
for each $k$ to determine $\hat{g_k}$. 
The second part is the update of the mixture proportions, given by
\begin{align}
	\hat{\pi_k} = \frac{1}{n}\sum_{i=1}^n P(Z_i=k\mid X_i).
\end{align}
A subtle change in the optimization objective explained in section \ref{altering} will enable us to calculate the MLE in the maximization step. Concretely, it suffices to change the first term of equation \eqref{altered_objective} into the normalized version of equation \eqref{weighted_likelihood}:
\begin{align}
	\sum_{i=1}^n \frac{P(Z_i=k\mid X_i)}{\sum_{j=1}^n P(Z_j=k\mid X_j)} \log g_k(X_i).
\end{align}
 The arguments thus far are still valid with this change.

To compare the results with other methods, we use the k-means++ algorithm \citep{vassilvitskii2006k} applied to this space as an alternative clustering method. Concretely, we implemented the usual k-means++ algorithm with the Fr\'{e}chet mean instead of with the usual arithmetic mean. It is simple to see that the sum of squared distances from the cluster centers strictly decreases at each iteration of the k-means++ or k-means algorithm  \citep{macqueen1967some}  as in the Euclidean case. Thus the algorithm terminates within a finite number of steps. 

\begin{figure}
\begin{center}
    \includegraphics[width=\linewidth]{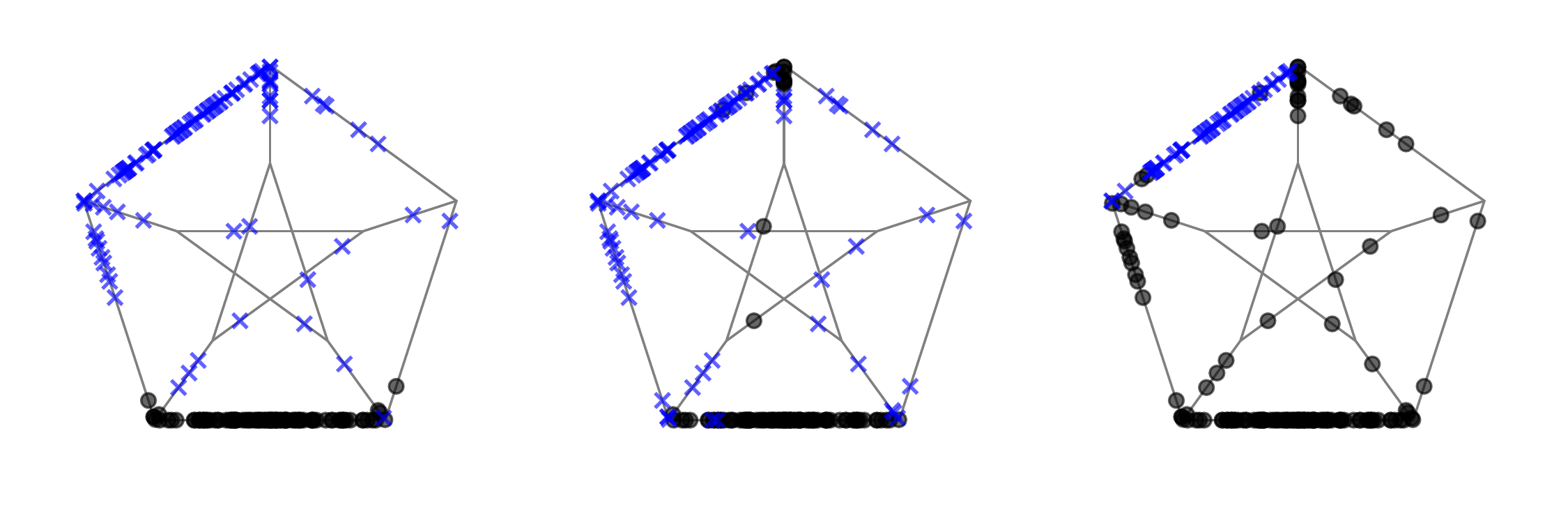}
\end{center}
	\caption{Clustered points displayed in the Petersen graph (Figure \ref{figure_treespace}) with labels. (Left): data points with markers indicating the true clusters to which they belong. (Center): clustering results using the log-concave mixture (89\% accuracy). (Right): clustering results using k-means++ (77 \% accuracy). Note that closeness in this graph does not necessarily mean closeness in $\mathcal{T}_4$.}
	\label{cluster_pp}
\end{figure}

The results of clustering are shown using the Petersen graph in Figure \ref{cluster_pp}. 
We see from these results that in this case, the clusters estimated by the log-concave mixture estimate are more representative of the true density. This is a natural result, as the true density is indeed log-concave, and the two mixed normal densities have different variances.

\section{Conclusion}
	In this paper, we showed that the maximum likelihood estimator exists in tree space under certain conditions. In one dimension, the maximum likelihood estimator exists with probability one and is unique, which is the same result as in the Euclidean case. In multiple-dimension, there are conditions under which the MLE does not exist. However, if we restrict our attention to a particular situation, we have shown that the MLE exists and is unique $\nu$-almost surely. Also, we have presented an algorithm to calculate the MLE exactly in the one-dimensional case and approximately in the two-dimensional case. We compared the results with the previously developed kernel density estimator and confirmed that our estimator dominates in well-modeled cases with a large enough sample size.
		
	The method derived here is promising in that it gives a new nonparametric approach to density estimation on tree space. As we saw in a two-dimensional example, it is also able to respect the support of the sample distribution when it is convex, which might affect the estimation accuracy considerably in some situations. Unlike kernel density estimates, we do not need either the determination of smoothing parameters or the calculation of any (approximate) normalizing constant. Although the computation is much slower than kernel density estimates, the accuracy of density estimation can be expected to be high when the log-concavity assumption is not far away from the properties of the true density.  The developed method gives an intermediate choice between fully unconstrained, nonparametric approaches, which have difficulties improving accuracy and estimation, and parametric approaches, for which it is difficult to specify the correct models for tree space.

	For future research, it is important to derive some theoretical properties of the estimator on tree space. In Euclidean space, log-concave MLE is known to be strictly consistent, and even if the model is misspecified, it converges to the ``log-concave projection'' of the true density (the log-concave density that minimizes Kullback-Leibler divergence from the true density). It is crucial to investigate whether these properties hold on tree space as well. Also, It is of interest whether we can introduce different constraints on the density in order to include other types of densities on this space, or to lessen the computational load. The density that bend at the boundary, which we considered only in the one-dimensional case, is a candidate class for this purpose. Finally, further simulation studies in both well-modeled and misspecified cases are necessary for practical purposes.
	
	As tree space is constructed for modeling the space of phylogenetic trees, one immediate interest of further research is the applicability to biological data and problems. As we only have (approximate) algorithms for one and two dimensions, corresponding to the case where a maximum of four taxa are present, the applicability seems to be limited for now. However, it might be possible to seek lower-dimensional representations of large trees, for example, by grouping some taxa when they are irrelevant in terms of the inconsistency of multiple trees. It is of interest how well our method with these modifications performs compared to the existing methods with the biological data. Further careful assessment of the log-concavity assumption about the data distribution is also of interest. Furthermore, an attempt to improve computational efficiency is called for, as our method currently is not suited for use with large datasets. 

\section*{Funding}
This work was supported by JSPS KAKENHI Grant Number JP22J22685.

\newpage

  \bigskip
  \bigskip
  \bigskip
  \begin{center}
    {\LARGE\bf Maximum Likelihood Estimation of Log-Concave Densities on Tree Space : Supplementary Material}
\end{center}
  \medskip

\appendix

\setcounter{figure}{8}    
\setcounter{theorem}{11}
\section{Proofs of Theorem 5 and Theorem 3}
In this section, we first derive some Lemmas related to the convex analysis in Hadamard spaces. Then we give a proof of Theorem 5, the existence of the maximum likelihood estimator in multi-dimensional cases. The proof of Theorem 3, the existence and uniqueness of the maximum likelihood estimator, follows similar arguments to the multi-dimensional cases (the proof of Theorem 5 and Theorem 4). Thus we only note the differences we need to consider.

\subsection{Lemmas}
First, we prepare some Lemmas. Let $(\mathcal{H},d)$ be a Hadamard space and for $i=1,2,\ldots,n$, $(X_i, y_i)\in \mathcal{H}\times\mathbb{R}$. Let $D_0 = \{X_1, \ldots, X_n\}$, and $C_n = \conv(X_1,\ldots,X_n)$. Lemma 1 suggests that this convex hull can be obtained by repeatedly taking geodesics starting from the set $D_0$. Concretely, $C_n = \cup_{l=0}^\infty D_l$, where $D_l$ consists of all points on the geodesics between points in $D_{l-1}$. We first show the boundedness of this convex hull.

\begin{lem}\label{lem:boundedness}
	Let $X_1, \ldots, X_n \in\mathcal{H}$. Then $C_n = \conv\{X_1, \ldots, X_n\}$ is a bounded set.  
\end{lem}

\begin{proof}[Proof of Lemma]
	This is proved using the nonpositive curvature property. By Lemma 1, \\$C_n = \cup_{l=0}^\infty D_l$, where $D_0 = \{X_1, \ldots, X_n\}$ and $D_l$ consists of all points that are on some geodesic with endpoints in $D_{l-1}$. Let $r = \max_{j=2,\ldots,n}d(X_1, X_j)$. We can recursively show that for any points $x_l \in D_l$, $d(X_1, x_l) \leq r$. To see this, first note that at $l=0$, the statement holds. Assuming that the statement holds at $l$, for any points $x_{l+1} \in D_{l+1}$, there exist $a_{l1}, a_{l2} \in D_l$ and $0\leq\mu\leq 1$ such that $x_{l+1} = \mu a_{l1} + (1-\mu) a_{l2}$. Then by CAT(0) inequality, $d(X_1, x_{l+1}) \leq \max_{k=1,2} d(X_1, a_{lk}) \leq r$. This shows the boundedness of $C_n$. 
\end{proof}

We say that $\mathcal{H}$ has the {\it geodesic extension property} if any geodesic between two points in $\mathcal{H}$ can be extended to a geodesic line. Now, we can show the following Lemma.

\begin{lem}\label{lem:continuous_interior}
    Suppose $\mathcal{H}$ has the geodesic extension property. Then,
    any upper-semicontinuous concave function $g$ on $\mathcal{H}$ that is bounded in the domain $S$ is continuous on $\mathrm{int}(S)$.
\end{lem}

\begin{proof}
    Let $g_{\mathrm{int}(S)}$ denote the restriction of $g$ to $\mathrm{int}(S)$. Then, the hypograph of this function is $\{ (x,\mu)\mid x\in \mathrm{int}(S), \mu\leq g(x) \}$. 
    
    We first show that the interior of this set is given by $T = \{ (x,\mu)\mid x\in \mathrm{int}(S), \mu< g(x) \}$. That $T$ includes $\mathrm{int}(\hypo g_{\mathrm{int}(S)})$ is obvious. We only need to show that $T$ is open. To see this, take any $(x,\mu)\in T$. Then there exists $\varepsilon_1 > 0$ such that $\mu + \varepsilon_1 = g(x)$. Because $x\in\mathrm{int}(S)$, we can take $\varepsilon_2>0$ such that the closed ball $\bar{B}_{\varepsilon_2}(x) = \{y\in \mathcal{H}\mid d(x,y)\leq \varepsilon_2\}$ is included in $\mathrm{int}(S)$. 
    Now, take $\varepsilon_3>0$ to satisfy $\varepsilon_2 > \varepsilon_3$. Then, by the geodesic extension property, we can see that for any $\tilde{y}\in B_{\varepsilon_3}(x)$, there exists $y\in\mathrm{int}(S)$ such that $d(x,y) = \varepsilon_2$ and for some $0<\theta<1$, $(1-\theta)x+\theta y = \tilde{y}$. Note that $\theta$ in the previous expression satisfies $\theta \leq \varepsilon_3/\varepsilon_2$. Because of concavity of $g$, $g(\tilde{y}) \geq g(x) - \theta(g(x)-g(y))) \geq g(x) - \varepsilon_3/\varepsilon_2 (g(x)-g(y))$. Since $g$ is bounded, by taking $\varepsilon_3$ sufficiently small, one can make for any $\tilde{y}\in B_{\varepsilon_3}(x)$, $g(\tilde{y}) > g(x) - \varepsilon_1/2$. Therefore, if we let $\varepsilon = \min\{\varepsilon_3, \varepsilon_1/2\}$, $B_{\varepsilon}(x)\subseteq T$. Thus, the set $T$ is open.
    
    For any $\alpha\in\mathbb{R}$, the intersection of two open sets $\mathrm{int}(\hypo(g_{\mathrm{int}(S)}))$ and $\{(x,\mu)\mid \mu>\alpha\}$ is an open set $\{(x,\mu)\mid x\in\mathrm{int}(C_n), g(x) > \mu >\alpha\}$, and thus the projection of this set $\{x\mid x\in\mathrm{int}(C_n), g(x) > \alpha\}$ . This implies the lower-semicontinuity, and thus the continuity, of $g$ on $\mathrm{int}(S)$.
\end{proof}

Now, denote by $\Delta^{n-1}$ the $(n-1)$-dimensional simplex in $\mathbb{R}^n$:
\begin{align}
    \Delta^{n-1} = \left\{\lambda = (\lambda_1,\ldots,\lambda_n)\in\mathbb{R}^n\mid \lambda_i\geq 0, \sum_{i=1}^n \lambda_i = 1\right\}
\end{align}
For $\lambda\in\Delta^{n-1}$ and $l=0,1,\ldots$, define sets $S_{\lambda, l}$ as follows:
\begin{align}
\begin{split}
    S_{\lambda,0} &= \begin{cases}
        \{X_i\} & \text{if }\lambda_i = 1, \lambda_j = 0 ~~(\forall j\neq i) \\
        \emptyset & \text{otherwise}
    \end{cases} \\
    S_{\lambda, l} &= \{ x\in D_l\mid \exists x_1\in S_{\lambda_1, l-1}, \exists x_2\in S_{\lambda_2, l-1}, 0 \leq \exists \mu\leq 1,\\&~~~~~~~ \gamma_{x_1,x_2}(\mu) = x, (1-\mu)\lambda_1 + \mu\lambda_2 = \lambda\}~~~~~~~~(l= 1,2,\ldots).
\end{split}
\end{align}
Let $S_{\lambda} = \cup_{l=0}^\infty S_{\lambda, l}$. Informally, $S_\lambda$ is the set of points which can be generated with cumulated coefficient of convex combination $\lambda$. Lemma 1 indicates that for any point $x\in C_n$, at least one $\lambda\in\Delta^{n-1}$ exists such that $x\in S_{\lambda}$. This also implies that $\conv(\{X_i,y_i\}_{i=1}^n)$ includes the point $(x, \lambda^\top y)$. 

Let $y=(y_1,\ldots,y_n)$ and denote by $f_y(x)$ the following function:
\begin{align}
    f_y(x) = \begin{cases}
        y_i & \text{if } x=X_i \\
        -\infty & \text{otherwise}.
    \end{cases}
\end{align}
Let $\bar{h}_y$ denote the upper-semicontinuous concave hull of $f_y$. Following two Lemmas show some properties of this function.

\begin{lem}
    $\bar{h}_y(x)$ is continuous with respect to $y = (y_1, \ldots, y_n)$.
    \label{lem:cont}
\end{lem}

\begin{proof}
    For any fixed value $x\in\mathcal{H}$, $\bar{h}_{y+\delta}(x)$ cannot deviate from $\bar{h}_y(x)$ by more than $\|\delta\|_{2}$.
    
    In order to see this, first, assume $\bar{h}_{y+\delta}(x)\leq\bar{h}_y(x)$. 
    By construction, the hypograph of $\bar{h}_y$ is given by
    \begin{align}
        \hypo~ \bar{h}_y = \cl (\conv (\hypo~f_y)) \label{eq:hypohy}
    \end{align}
    Let $\mu=\bar{h}_y(x)$. Equation \eqref{eq:hypohy} indicates that there exists $\{(x_i, \mu_i)\}_{i=1}^\infty \subseteq\conv(\hypo~f_y)$ such that the sequence $\{(x_i, \mu_i)\}_{i=1}^\infty$ converges to $(x, \mu)$. Furthermore, by Lemma 1, there exists $\lambda^{(i)}\in\Delta^{n-1}$ for each $(x_i, \mu_i)$ such that $x_i\in S_{\lambda^{(i)}}$ and $\mu_i \leq (\lambda^{(i)})^\top y$. This implies that $(x_i, (\lambda^{(i)})^\top (y+\delta))\in \hypo(\bar{h}_{y+\delta})$, $\mu_i - \|\delta\|_{2} \leq (\lambda^{(i)})^\top (y+\delta)$, and thus $(x_i, \mu_i - \|\delta\|_{2}) \in \hypo(\bar{h}_{y+\delta})$. Now, the sequence of points $\{(x_i, \mu_i - \|\delta\|_{2})\}_{i=1}^\infty$ converges to $(x, \mu-\delta)$, and this points is included in $\hypo(\bar{h}_{y+\delta})$ by the upper-semicontinuity. This shows that $\bar{h}_{y+\delta}(x)\geq \mu-\|\delta\|_{2}$ and consequently, $\bar{h}_y(x) - \bar{h}_{y+\delta}(x) \leq \|\delta\|_2$. 
    
    If $\bar{h}_{y+\delta}(x)>\bar{h}_y(x)$, let $\mu=h_{y+\delta}(x)$, and by interchanging the roles of $\bar{h}_y$ and $\bar{h}_{y+\delta}$ in the above argument, we can show $\bar{h}_{y+\delta}(x) - \bar{h}_{y}(x) \leq \|\delta\|_2$. 
\end{proof}

\begin{lem}\label{lem:domain_closure}
    $\mathrm{dom}~\bar{h}_y = \cl(C_n)$.
\end{lem}
\begin{proof}
    For any upper-semicontinuous concave function $g$ satisfying $g(X_i)\geq y_i$, let the restriction of $g$ to $\cl(C_n)$ be $g_{\cl(C_n)}$. Then $\hypo~g_{\cl(C_n)} = \hypo~g \cap \{(x,\mu)\mid x\in\cl(C_n), \mu\in\mathbb{R}\}$ is closed. This means that $g_{\cl(C_n)}$ is upper-semicontinuous. Furthermore, since $X_i\in C_n$, $g_{\cl(C_n)}(X_i) = g(X_i)\geq y_i$. With the minimality of $\bar{h}_y$, this implies that $\dom ~\bar{h}_y\subseteq \cl(C_n)$.
    
    On the other hand, take any $x\in \cl(C_n)$. Then there exists a sequence $\{x_i\}_{i=1}^\infty \subseteq C_n$ such that $x = \lim_{i\to\infty} x_i$. By the concavity of $\bar{h}_y$, $\bar{h}_y(x_i)\geq \min_{j=1,2,\ldots,n} y_j$ for all $i=1,2,\ldots$. By the upper-semicontinuity, $\min_{j=1,2,\ldots,n} y_j \leq \lim_{i\to\infty} f(x_i)\leq f(x)$. This shows that $x\in\mathrm{dom}~\bar{h}_y$, and consequently, $\cl(C_n)\subseteq \mathrm{dom}~\bar{h}_y$.
\end{proof}

\subsection{Proof of Theorem 5}
\label{multidimproof}

\begin{proof}[Proof of theorem]
    The proof is essentially a modification of the proof of Theorem 1 from \citet{Cule2010b}.
    
	By Lemma 1, $C_n = \cup_{l=0}^\infty D_l$, where $D_l$ denotes the set defined in the proof of the previous Lemma. Let $\bar{\mathcal{F}}$ be the set of upper-semicontinuous log-concave functions on $\mathcal{T}_{p+2}$, and consider the maximization of the function $\psi_n(f) = n^{-1}\sum_{i=1}^n \log f(X_i) - \int_{\mathcal{T}_{p+2}}f(x)d\nu(x)$ over $\bar{\mathcal{F}}$.

First, we can show that for any $g\in \bar{\mathcal{F}}$, there exists $f\in \bar{\mathcal{F}}$ such that $\psi_n(f)\geq \psi_n(g)$ with 
\begin{align}
f(x) > 0~(x\in \cl(C_n)), ~~f(x) = 0~(x\not\in \cl(C_n)). \label{eq:cond2}
\end{align}
First, assume $g(x)=0$ for some $x\in C_n$. Lemma 1 suggests that there exists $l\in\mathbb{N}$ such that $x\in D_l$.  If $l=1$, we have for some $k$, $g(X_k)=0$, implying $\psi_n(g) = -\infty$. Otherwise, take minimum such $l$, then one can take some $x_{1, l-1}, x_{2, l-1} \in D_{l-1}$ and $\mu\in(0,1)$ such that $x = (1-\mu)x_{1,l-1} + \mu x_{2,l-1}$. By concavity, $-\infty = \log g(x) \geq (1-\mu)\log g(x_{1, l-1})+ \mu\log g( x_{2, l-1})$. This leads to that at least one of $g(x_{1,l-1})$ and $g(x_{1,l-2})$ must be zero. Repeating this process, we can show that there exists $k\in\{1,\ldots, n\}$ such that $g(X_k) = 0$, resulting again in $\psi_n(g) = -\infty$. Therefore, if for some $x\in C_n, g(x) = 0$, then any $f$ with \eqref{eq:cond2} satisfies $\psi_n(f) \geq \psi(g)$. 
If $g(x)=0$ for some $x\in\cl(C_n)\backslash C_n$, then one can take a sequence $\{x_i\}\in C_n$ such that $x_i$ converges to $x$. The upper-semicontinuity implies that $\limsup_{i\to\infty} g(x_i)\leq g(x) = 0$. Then for all $\varepsilon>0$, there exists $i$ such that $g(x_i)<\varepsilon$. This with the previous argument leads to that $\min_{k=1,\ldots,n}g(X_k) < \epsilon$. Since $\epsilon$ can be taken arbitrarily small, $\min_{k=1,\ldots,n}g(X_k) = 0$, which leads again to $\psi_n(g) = -\infty$. 
On the other hand, because $\cl(C_n)$ is bounded by Lemma \ref{lem:boundedness}, if we take $f\in\mathcal{F}$ to be finitely positive and upper-semicontinuous on $\cl(C_n)$, $f$ becomes  upper-semicontinuous on $\dom~f$ and $\psi_n(f) > -\infty$. Thus, we can strictly restrict our attention to $f$ such that $f(x)>0$ for all $x\in C_n$.

Now, assume that $g(x) > 0$ for some $x\not\in \cl(C_n)$. Then by letting $f(x)$ be the restriction of $g(x)$ to $\cl(C_n)$, it is easy to see that $f$ satisfies \eqref{eq:cond2}, log-concavity, and upper-semicontinuity. Furthermore, $\psi_n(f) \geq \psi_n(g)$, and the inequality becomes the strict one if $\nu(\{x~|~x\in\dom~ g, x\not\in C_n \}) > 0$. This means that we can strictly restrict attention to the density that satisfies \eqref{eq:cond2}, or the ones that satisfy them except for some set of $\nu-$measure zero excluding $\cl(C_n)$.

Secondly, we can show that for any upper-semicontinous function $g$, there exists $f$ which has its logarithm in the form $\bar{h_y}$ such that $\psi_n(f)\geq\psi_n(g)$. This is achieved if we put $y_i = \log g(X_i)$, $y=(y_1,\ldots,y_n)$ and $f=\exp(\bar{h_y})$. 
Note that by Lemma \ref{lem:domain_closure}, $\mathrm{dom}~\bar{h_y} = \cl(C_n)$. Also, by Lemma \ref{lem:continuous_interior}, $\bar{h_y}$ is continuous relative to $\mathrm{int}(C_n)$. Because of continuity, if $g(x)\neq \exp(\bar{h_y})(x)$ at some $x\in\mathrm{int}(C_n)$, $\psi_n(\exp(\bar{h_y}))>\psi_n(g)$. This shows that a maximizer of $\psi_n$, if it exists, has to take the form $\psi_n(\bar{h_y})$ in $\mathrm{int}(C_n)$.

Next, we can also show that we can restrict attention to the case $f\in \mathcal{F}_0$. For any upper-semicontinous function $g$, assume $\int_{\mathcal{T}_{p+2}} g(x) = c < \infty$. By setting $f=g/c$, $f\in\mathcal{F}_0$, 
\begin{align}
    \psi_n(f) &= \frac{1}{n}\sum_{i=1}^n \log f(X_i) - \int_{T_{p+2}}f(x)d\nu(x) \nonumber \\
	&= \frac{1}{n}\sum_{i=1}^n \log g(X_i) -\log c - 1 \nonumber \\
	&= \psi_n(g) + c -\log c -1 \nonumber  \\
	&\geq \psi_n(g).
\end{align}
Equality is attained only when $c=1$.

For the existence of a maximizer of $l(f)$, it only remains to show that $\psi_n$ has a maximizer $f = \exp(\bar{h_y})\in \mathcal{F}_0$ with \eqref{eq:cond2}. 

This is seen as follows. For an arbitrary function $f=\exp(\bar{h_y})$ with \eqref{eq:cond2}, let $X_{\max} = \argmax_{X_i} g(X_i)$ and $O_{\max}$ be one of the nonnegative orthants 
of $\mathcal{T}_{p+2}$ which $X_{\max}$ belongs to. Then by assumption, $\nu(C_n\cap O_{\max}) \eqqcolon V > 0$. Now, put $M=\bar{h_y}(X_{\max})$ and $m=\min_{x\in O_{\max}}\bar{h_y}(x)$. For sufficiently large $M$, $M-1>m$. Then by setting $\lambda=1/(M-m)$, the concavity $\bar{h_y}(\lambda x + (1-\lambda)X_{\max}) \geq \lambda\bar{h_y}(x) + (1-\lambda) \bar{h_y}(x)$ indicates 
\[\bar{h_y}\left(\frac{1}{M-m}x + \frac{M-m-1}{M-m}X_{\max}\right) \geq \frac{m}{M-m} + \frac{(M-m-1)M}{(M-m)} = M-1.\] 
This in turn indicates $\nu(\{x\mid\bar{h_y}(x) \geq M-1\})\geq V/(M-m)^p$, and thus $1 = \int_{\mathcal{T}_{p+2}} \exp(\bar{h_y}(x))d\nu(x) \geq V(\exp(M-1))/(M-m)^p$. Therefore,
\begin{align}
	M - m &\geq V^{\frac{1}{p}}(\exp(M-1))^{\frac{1}{p}} \nonumber \\
	\therefore m &\leq M - V^{\frac{1}{p}}\exp\left(\frac{M-1}{p}\right).
\end{align}
This leads to the following inequality:
\begin{align}
	\psi_n(\exp(\bar{h_y})) &= \frac{1}{n}\sum_{i=1}^n \bar{h_y}(X_i) - 1 \nonumber \\
	&\leq \frac{n-1}{n} M + \frac{M}{n} - \frac{V^{\frac{1}{p}}}{n}\exp\left(\frac{M-1}{p}\right) \nonumber \\
	&= M - \frac{V^{\frac{1}{p}}}{n}\exp\left(\frac{M-1}{p}\right)
\end{align}
This inequality shows that as $M$ goes to infinity, $\psi_n$ goes to $-\infty$. 
On the other hand, because $\bar{h_y}$ is continuous with respect to $y$, as shown in Lemma \ref{lem:cont}, $\psi_n(\exp(\bar{h_y})) = n^{-1}\sum_{i=1}^n \bar{h_y}(X_i) - \int_{\mathcal{T}_{p+2}} \exp(\bar{h_y})(x)dx$ is also continuous. Thus, there exists $r = \max_{y\in[-M, M]^n} \psi_n(\exp(\bar{h_y}))$. This shows the existence of a maximizer of $\psi_n(f)$, and thus $l(f)$.

The uniqueness of the maximizer in the $\nu$-almost sure sense is guaranteed by Theorem 4. Moreover, by the previous argument, the maximizer needs to be the form $\exp(\bar{h}_y)$ for some $y\in\mathbb{R}^d$, in $\mathrm{int}(C_n)$. Since $\bar{h_y}$ is continuous in $\mathrm{int}(C_n)$, the strict uniqueness in $\mathrm{int}(C_n)$ holds.
\end{proof}

\subsection{Proof of Theorem 3}
\label{onedimproof}
\begin{proof}
    The proof goes in a similar way as in the proof of Theorem 5. Here, we note several properties specific to the one-dimensional case, which is necessary for the proof.
    \begin{itemize}
        \item $C_n = \mathrm{conv}(X_1, \ldots, X_n)$ is a closed bounded set. This, in particular, implies $\cl(C_n) = C_n$. Thus the domain of the maximizer can be restricted to exactly $C_n$.
        \item The logarithm of the maximizer can always be written in the form $h_y$, the least concave function with $h_y(X_i)\geq y_i$. $h_y$ is continuous on $C_n$.
        \item Because $f$ is absolutely continuous with respect to $\nu$, with probability 1, $C_n$ is not merely a point, but constitutes an interval. This is essentially a substitute for the sufficient condition (a)-(c) we considered in the multi-dimensional case. We can easily see that this condition is sufficient for showing that the maximizer of $\psi_n$ exists in the set of densities of the form $\exp(h_y)$.
        \item The strict uniqueness follows from the following two observations: $h_y$ is continuous on $C_n$, and the extension of the domain always provokes the increase in the measure.
    \end{itemize}

\end{proof}

\section{Derivation of distribution of $3$-trees under coalescent}
In this section, we derive the equation (13) in Example 8, the distribution of $3$-trees generated from multi-species coalescent. 

In order to consider the distribution on tree space, we need to consider the distribution of gene tree topology and the internal edge length given species tree. Algorithms for calculating gene tree topology given specific species trees have been considered in general dimension \citep{Degnan2005, Wu2012}, but the direct calculation is also available in small trees \citep[e.g.][]{Takahata1985, Pamilo1988}. In the case of simplest case of $3$-trees, denote by $u_1$ is the topology of the species tree and by $u_2, u_3$ the rest of the topologies. Then the distribution of gene tree topology $U_g$ is given by:
\begin{align}
\begin{split}
    P\left(U_g = u_1\right) &= 1-\frac{2}{3}e^{-T}, \\
    P\left(U_g = u_j\right) &= \frac{1}{3}e^{-T} ~~~~~~~ (j=2,3),
\end{split}\label{eq_congruence}
\end{align}
where $T$ denotes the length of the internal edge of the species tree.

Given $U_g$, the distribution of the internal edge length can be derived in the following manner. 
First, consider the cases where $U_g \neq u_1$. In these cases, the distribution of the internal edge length is the same as the distribution of the coalescence time of two lineages. Standard coalescent theory implies that the density of the internal edge length is given by
\begin{align}
    f(x~|~U_g = u_j) = \exp({-x}).
\end{align}

\begin{figure}
    \centering
    \includegraphics[width=.4\linewidth]{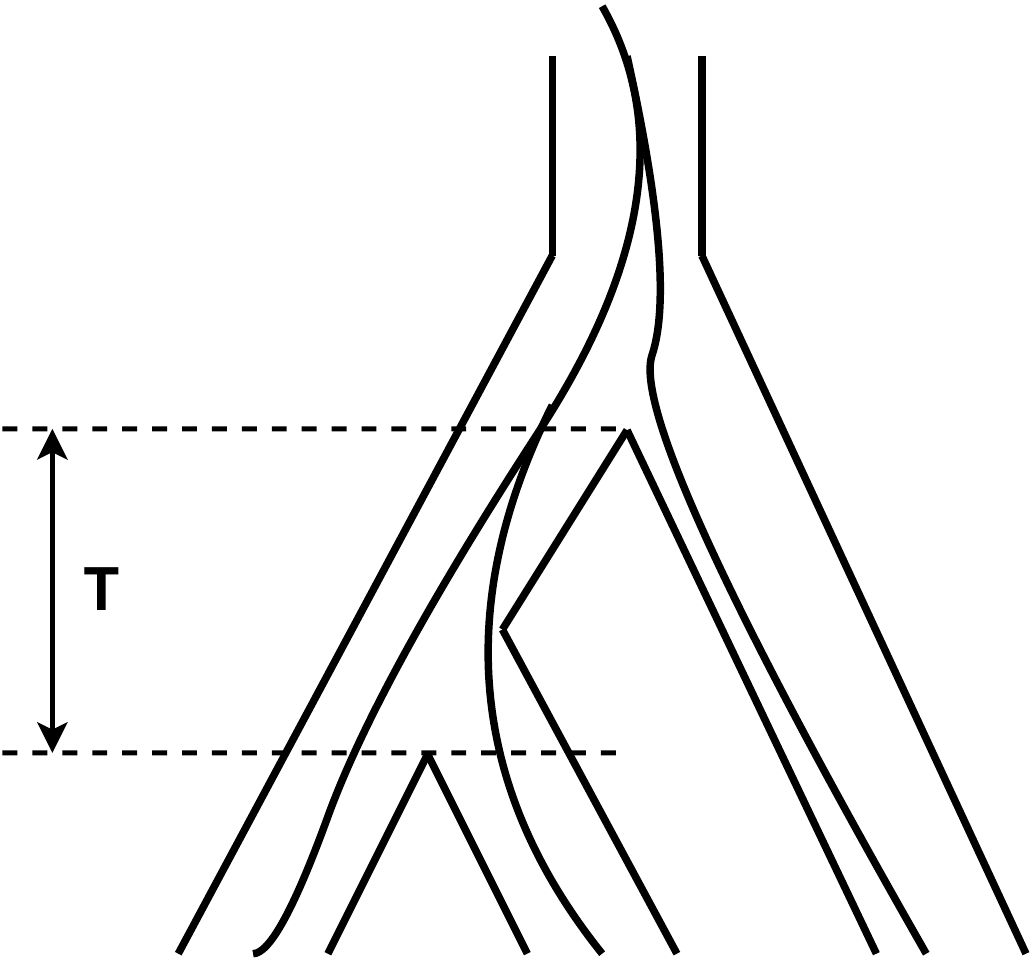}
    \caption{Illustration of coalescence of three lineages from three different species. $\tau$ denotes the time of the first coalescence of genes after the first coalescence of the species tree.}
    \label{fig:tau}
\end{figure}

Next, we consider the case $U_g = u_1$. Let $\tau$ denote the time of the first coalescence of any two lineages after the first coalescence of the species tree coalescence (Figure \ref{fig:tau}). Thus, the value of $\tau$ being close to zero means that the two gene lineages coalesce right after the species tree coalesces, and the value of $\tau$ larger than $T$ means that two gene lineages coalesce after whole three species coalesced in the species tree. Multi-species coalescent assumes that the coalescence is neutral. In terms of $\tau$, this can be written as 
\begin{align}
    P(U_g = u_1\mid \tau) = 
    \begin{cases}
        1 & \text{if } \tau \leq T \\
        \frac{1}{3} & \text{otherwise}
    \end{cases} \label{tg}.
\end{align}

When $\tau\leq T$, the density of $\tau$ is $\exp({-\tau})$. This, combined with the equation \eqref{tg}, implies that the conditional density of $\tau$ given $U_g=u_1$ for $\tau\leq T$ is given by
\begin{align}
    f(\tau\mid U_g &= u_1) = \frac{1}{1 - (2/3)\exp(-T)}\exp(-\tau) & \text{if } \tau\leq T.
\end{align}

We can now calculate the density of the internal edge length $x$ given $U_g = u_1$ as follows:

\begin{align}
    &f(x\mid U_g = u_1) \nonumber\\
    =& \int_0^T f(x\mid U_g=u_1, \tau) f(\tau\mid U_g=u_1)d\tau + \int_T^\infty f(x\mid U_g=u_1, \tau) f(\tau\mid U_g=u_1)d\tau \nonumber\\
    =& \left(1-\frac{2}{3}\exp(-T)\right)^{-1} \int_0^T \exp(-(x-(T-\tau))) \mathbbm{1}(x\geq T-\tau) \exp(-\tau)d\tau \nonumber\\
    &+ \int_T^\infty \exp(-x) f(\tau\mid U_g = u_1)d\tau \nonumber \\
    =& \left(1-\frac{2}{3}\exp(-T)\right)^{-1} \int_{\max\{0, T-x\}}^T \exp(- x + T) \exp(-2\tau)d\tau \nonumber\\
    &+ \exp(-x)\left\{ 1 - \int_{0}^T f(\tau\mid U_g = u_1)d\tau \right\}\nonumber\\
    =& \left(1-\frac{2}{3}\exp(-T)\right)^{-1}\left\{-\frac{1}{2}\exp(-x+T)\left[ \exp(-2\tau) \right]_{\max\{0, T-x\}}^T \right\} \nonumber\\
    &+ \frac{\exp(-T)}{3}\left(1-\frac{2}{3}\exp(-T)\right)^{-1} \exp(-x) \nonumber\\
    &= \left(1-\frac{2}{3}\exp(-T)\right)^{-1}\left[-\frac{1}{6}\exp(-x-T) + \frac{1}{2}\exp(-x+T-2\max\{0,T-x\})\right] \label{eq_cond_ie}
\end{align}

Equations \eqref{eq_cond_ie}, \eqref{eq_congruence} leads to the equation (13) in Example 8.

\section{Bounds for maximum and minimum values at the origin generated by geodesics in $H_{l-1}$}
As mentioned in the main text, finding the maximum and minimum values at the origin in the first step of this approximation algorithm is not a simple optimization problem. For example, the problem of finding the maximum value at the origin among all geodesics between points in two specific orthants, say $O_1$ and $O_2$, can be written in the following way: given points $\{x_i, y_i\}\in O_1\times\mathbb{R}$ and $\{p_j, q_j\}\in O_2\times\mathbb{R}$, 
\begin{equation}
\begin{aligned}\label{optim_prob_eg1}
    \max_{{\mu_i}, {\lambda_j}} &~~~ \frac{\|\sum_i \mu_i x_i\|\sum_{j} \lambda_j q_j + \|\sum_j \lambda_j p_j\|\sum_i\mu_i y_i}{\|\sum_i \mu_i x_i\| + \|\sum_j \lambda_j p_j\|} \\
    \text{s.t.} &~~~ \sum_{i}\mu_i = 1 \\
    &~~~ \sum_j \lambda_j = 1 \\
    &~~~ 1\geq \mu_i, \lambda_j \geq 0 \\
     &~~~ \text{the geodesic between } \sum_{i}\mu_ix_i\text{ and }\sum_j\lambda_jp_j\text{ is a cone path.}
\end{aligned}
\end{equation}
Although the last nonconvex constraint can be removed when we are considering some combinations of two orthants, the objective function is also nonconvex. The next is an example that shows nonconvexity in a very simple case.

\begin{example}\label{example:opt}
    Consider two orthants $O_1$ and $O_2$ separated by two other orthants so that the geodesic between points in two orthants are necessarily cone paths. We assume that we have two points in each orthant, say $x_1\in O_1, x_2\in O_1, p_1\in O_2, p_2\in O_2$, with the coordinates at each orthant being $(0,1), (1,0), (0,1), (1,0)$ respectively. We consider the problem \eqref{optim_prob_eg1} with four points $(x_1,1), (x_2, 1), (p_1, -1), (p_2, -1)$ (Figure \ref{kantanopt}). 
 \begin{figure}
    \begin{minipage}{0.33\linewidth}
	\begin{center}
		\includegraphics[width=\linewidth]{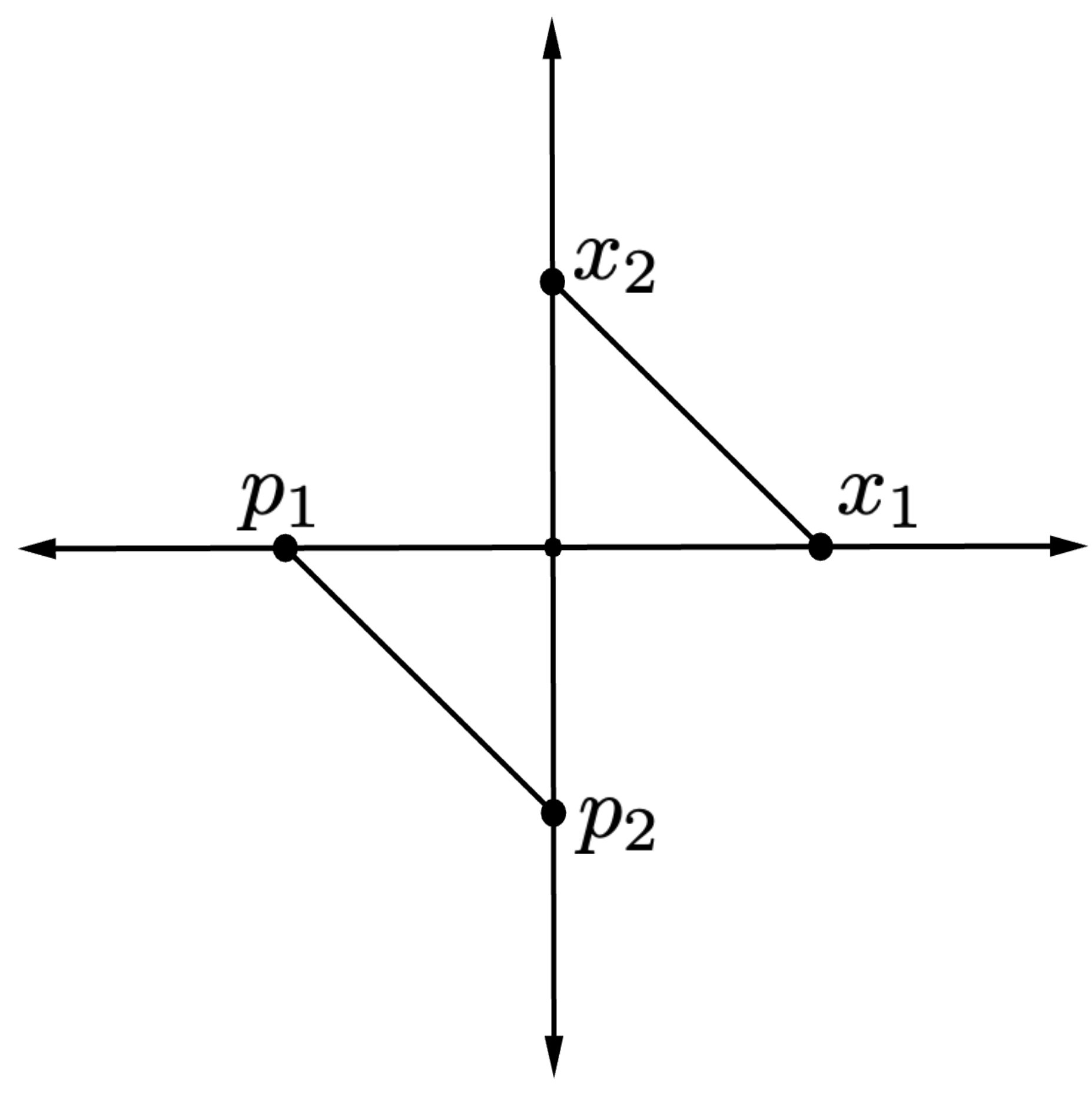}
	\end{center}
	\end{minipage}
	\begin{minipage}{0.31\linewidth}
	\begin{center}
		\includegraphics[width=.9\linewidth]{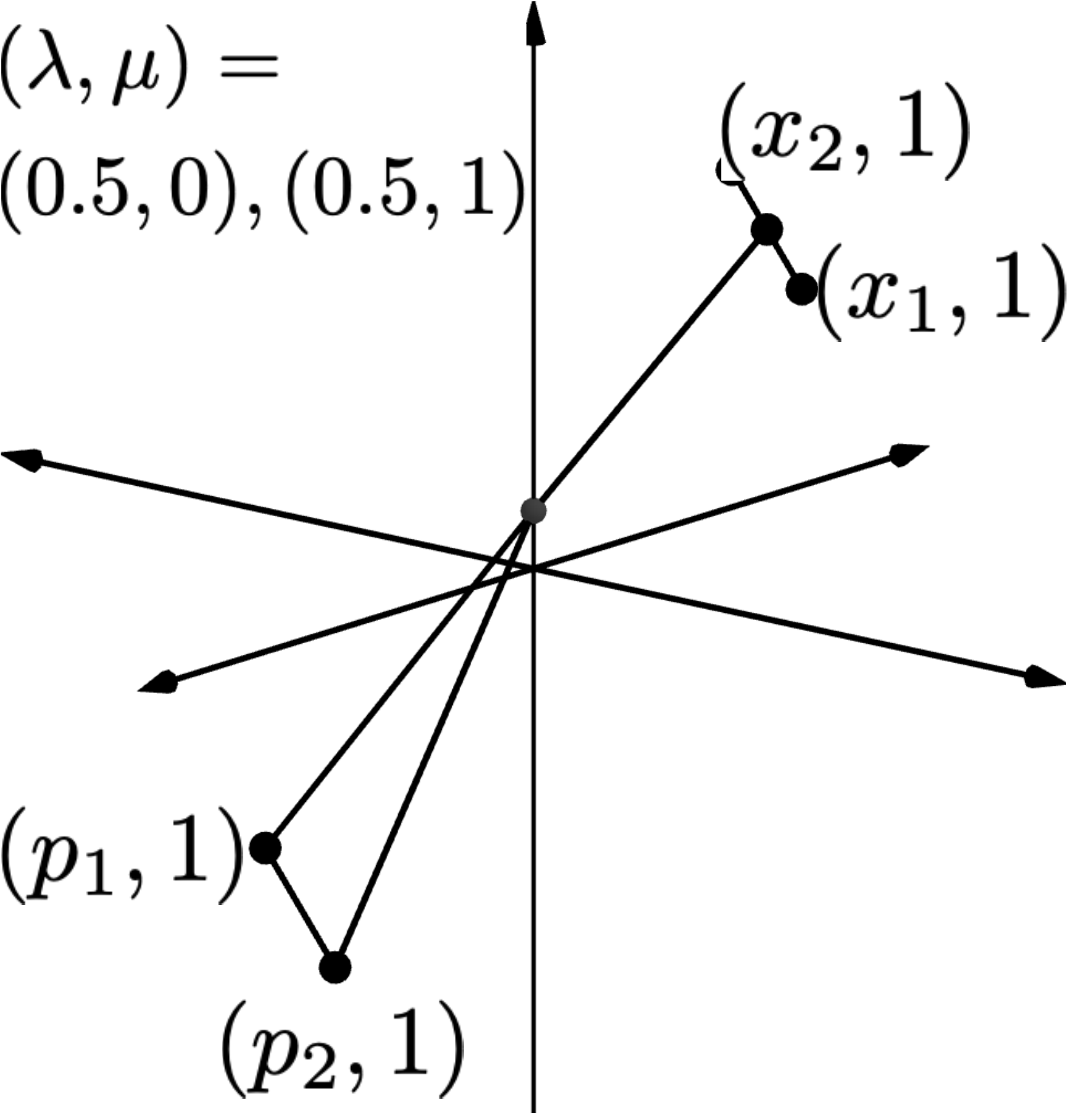}
	\end{center}
	\end{minipage}
	\begin{minipage}{0.31\linewidth}
	\begin{center}
		\includegraphics[width=.9\linewidth]{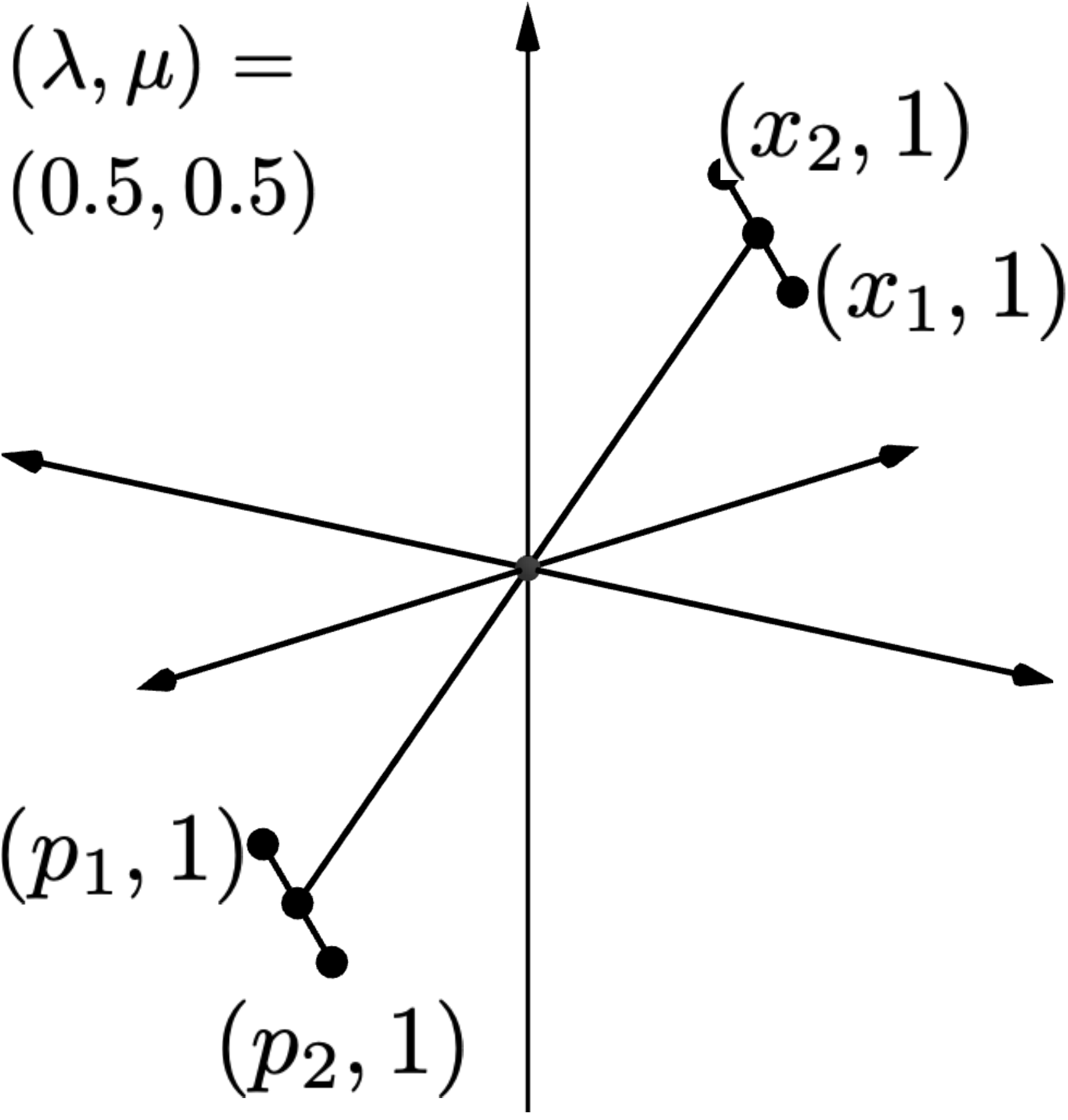}
	\end{center}
	\end{minipage}
	\caption{Points used in Example \ref{example:opt}. (Left): four points $x_1, x_2, p_1, p_2$. (Center): the optimal cone paths with $(\lambda, \mu) = (0.5, 1), (0.5,0)$. (Right): the cone path when $(\lambda, \mu) = (0.5, 0.5)$.}
	 \label{kantanopt}
\end{figure}
    
    Then the optimization problem to solve is
    \begin{equation}
    \begin{aligned}\label{optim_prob_eg2}
    \max_{\mu, \lambda} &~~~ f(\mu, \lambda) = \frac{-\sqrt{\mu^2 + (1-\mu)^2} + \sqrt{\lambda^2 + (1-\lambda)^2}}{\sqrt{\mu^2 + (1-\mu)^2} + \sqrt{\lambda^2 + (1-\lambda)^2}} \\
    \text{s.t.} &~~~ 1 \geq \mu, \lambda \geq 0 
\end{aligned}
\end{equation}
The optimal solutions to this problem are $(\lambda, \mu) = (0.5, 1)$ and $(\lambda, \mu) = (0.5, 0)$. Moreover, $(\lambda, \mu) = (0.5, 0.5)$ is a saddle point, and the second derivatives are
\begin{align}
    \frac{\partial^2 f}{\partial\mu^2}\mid_{\mu=0.5, \lambda=0.5} &= -2, \\
     \frac{\partial^2 f}{\partial\lambda^2}\mid_{\mu=0.5, \lambda=0.5} &= 2.
\end{align}
Thus, the objective function is concave in $\mu$ and convex in $\lambda$ around $(\mu, \lambda)=(0.5, 0.5)$.
\end{example}

As it is computationally intensive to solve this nonconvex optimization problem at each iteration, we replace the maximum (resp. minimum) values at the origin with its lower bound (resp. upper bound), which can be easily computed by linear programming. In the following paragraph, we explain the algorithm to calculate the replacement value.

First, note that in the two-dimensional tree space, there are 15 orthants, each consisting of two axes at a right angle, and each axis is connected to two other orthants. Thus, there are $15\times 4 = 60$ combinations of three orthants connected sequentially, as one can see by Figure 1 in the main text. We denote these combinations of orthants as $\{(O_{k1}, O_{k2}, O_{k3})\}_{k=1}^{60}$, where $O_{k1}, O_{k2}, O_{k3}$ are connected in this order.

Figure 4 in the main text is a valid representation of orthants $O_{k1}, O_{k2}, O_{k3}$ in the Euclidean space, in the sense that the distance between two points in these orthants is respected if we do not allow the paths to cross the space outside these three orthants, which is the hatched region on the left in Figure 4. Then, the space $\cup_{j=1}^3 O_{kj} \times \mathbb{R}$ constitutes a part of $\mathbb{R}^3$, as depicted on the right in Figure 4. For all 60 combinations of these orthants,  we embed points of $S_{l-1}$ in $\mathbb{R}^3$, calculate the usual three-dimensional convex hull in this space (allowing paths to cross the quadrant which is not used for the embedding), and if the convex hull includes a point of the form $\{(0,y)\}$, where $0$ is the origin, then set the maximum and minimum $y$-values at the origin as $y_{l,k1}$ and $y_{l,k0}$, respectively. Note that the convex hull calculated here includes points outside the embeddings. In practice, we do not need to compute the whole convex hull. Rather, we only need to solve a linear programming problem. In the case of finding maximum values, given Euclidean embeddings $\{(x_{km}, y_{km})\}_{m=1}^{n_k}$, the problem is written as follows:
\begin{equation}
    \begin{aligned}\label{optim_prob_eg3}
    \max_{\{\lambda_m\}_{m=1}^{n_k}} &~~~ \sum_{m=1}^{n_k} \lambda_m y_{km}, \\
    \text{s.t.} &~~~ \sum_{m=1}^{n_k} \lambda_m = 1, \\
    &~~~1 \geq \lambda_m \geq 0 .
\end{aligned}
\end{equation}
Then let the maximum of all $\{y_{l,k1}\}$ and the minimum of all $\{y_{l,k0}\}$ be $y_{l1}^\prime$ and $y_{l0}^\prime$, respectively, and make them replacements for $y_{l1}$ and $y_{l0}$. 

Note that $y_{l1}^\prime$ (resp. $y_{l0}^\prime$) is a valid lower (resp. upper) bound for $y_{l1}$ (resp. $y_{l0}$), since there always exists a cone path geodesic between points in $S_{l-1}$ that crosses $(0,y_{l1}^\prime)$ and $(0,y_{l0}^\prime)$. Also, as we can see from the proof of Theorem 10, this procedure assures that all geodesics between points in $H_{l-1}$ that are not cone paths are always included in $H_l$.

\bibliography{LCDTreeSpacePaper}
\end{document}